\documentclass[pdflatex,article,12pt]{article}
\usepackage{amsmath}
\usepackage{graphicx}
\usepackage{enumerate}
\usepackage[round,authoryear]{natbib}
\usepackage{url} 
\usepackage{verbatim}
\usepackage[title]{appendix}%
\usepackage{algorithm}
\usepackage{algpseudocode}
\usepackage{xcolor}
\usepackage{rotating}
\usepackage{subcaption}  
\usepackage{caption}  

\usepackage{ifthen}

\usepackage{hyperref}
\hypersetup{
    colorlinks=true,
    linkcolor=blue, 
    citecolor=blue,
    urlcolor=blue
}

\newcommand{\blind}{0}

\newcommand{\rpack}{\if0\blind
	{\url{github.com/mkln/bipps}}\fi
	\if1\blind
	{\url{github.com/}\texttt{[url redacted in blinded version]}}\fi}

\newcommand{\codereproduce}{\if0\blind
	{\url{github.com/jeliason/bipps_paper_code}}\fi
	\if1\blind
	{\url{github.com/}\texttt{[url redacted in blinded version]}}\fi}

\usepackage{amsmath, amsthm, amssymb, amsfonts, bbm, setspace, bigints}%
\usepackage{graphicx}%
\usepackage{booktabs}%
\usepackage{enumerate}%
\usepackage{ragged2e}%
\usepackage{csquotes}%
\usepackage{textcomp}%
\usepackage{hhline}%
\usepackage{multirow}%
\usepackage{xcolor,colortbl}%
\usepackage{url}%
\usepackage{chngcntr}%
\usepackage{lipsum}%
\usepackage{lmodern}%
\usepackage{tcolorbox}%
\usepackage{comment}%
\usepackage{rotating}%
\usepackage{lscape}%
\usepackage{soul}%
\usepackage{listings}%

\renewcommand{\tilde}[1]{\widetilde{#1}}

\newcommand{\bolds}[1]{\boldsymbol{#1}}

\newcommand{\calD}{{\cal D}}

\newcommand{\calW}{{\cal W}}
\newcommand{\calY}{{\cal Y}}
\newcommand{\Cov}{\bolds{C}}

\newcommand{\bA}{\bolds{A}}

\newcommand{\bB}{\bolds{B}}

\newcommand{\bF}{\bolds{F}}

\newcommand{\bh}{\bolds{h}}

\newcommand{\bl}{\bolds{\ell}}

\newcommand{\bv}{\bolds{v}}
\newcommand{\bV}{\bolds{V}}
\newcommand{\bw}{\bolds{w}}

\newcommand{\bx}{\bolds{x}}
\newcommand{\bX}{\bolds{X}}
\newcommand{\by}{\bolds{y}}

\newcommand{\bzero}{\mathbf{0}}

\newcommand{\balpha}{\bolds{\alpha}}
\newcommand{\bbeta}{\bolds{\beta}}

\newcommand{\btheta}{\bolds{\theta}}

\newcommand{\blambda}{\bolds{\lambda}}

\newcommand{\bvarphi}{\bolds{\varphi}}

\usepackage{mathtools}

\newcommand{\be} {\begin{eqnarray*}}
\newcommand{\ee} {\end{eqnarray*}}

\definecolor{Gray}{gray}{0.85}
\definecolor{LightCyan}{rgb}{0.88,1,1}

\newcolumntype{a}{>{\columncolor{Gray}}c}

\renewcommand{\baselinestretch}{1.20}
\newcommand{\customfootnotetext}[2]{{
  \renewcommand{\thefootnote}{#1}
  \footnotetext[0]{#2}}}

\begin{document}

\def\spacingset#1{\renewcommand{\baselinestretch}%
{#1}\small\normalsize} 

\if0\blind
{
 \spacingset{1}
  \title{\bf Joint Modeling of Spatial Dependencies Across Multiple Subjects in Multiplexed Tissue Imaging}
  \date{}
  
  \author{%
  Joel Eliason\textsuperscript{a} 
  \and Arvind Rao \textsuperscript{a,b,c}
  \and Timothy L. Frankel \textsuperscript{d}
  \and Michele Peruzzi \textsuperscript{c,$\star$} 
  }

  \customfootnotetext{a }{Department of Computational Medicine \& Bioinformatics, University of Michigan--Ann Arbor} 
  \customfootnotetext{b }{Department of Radiation Oncology, University of Michigan--Ann Arbor}
  \customfootnotetext{c }{Department of Biostatistics, University of Michigan--Ann Arbor}
  \customfootnotetext{d }{Department of Surgery, University of Michigan--Ann Arbor}
  \customfootnotetext{$\star$ }{Corresponding author: \url{peruzzi@umich.edu}.}
  \maketitle
} \fi

\if1\blind
{
\spacingset{2}
  \bigskip
  \bigskip
  \bigskip
  \begin{center}
    {\LARGE\bf Joint Modeling of Spatial Dependencies Across Multiple Subjects in Multiplexed Tissue Imaging}
\end{center}
  \medskip
} \fi

\bigskip

\spacingset{1}

\begin{abstract}
The tumor microenvironment (TME) is a spatially heterogeneous ecosystem where cellular interactions shape tumor progression and response to therapy. Multiplexed imaging technologies enable high-resolution spatial characterization of the TME, yet statistical methods for analyzing multi-subject spatial tissue data remain limited. We propose a Bayesian hierarchical model for inferring spatial dependencies in multiplexed imaging datasets across multiple subjects. Our model represents the TME as a multivariate log-Gaussian Cox process, where spatial intensity functions of different cell types are governed by a latent multivariate Gaussian process. By pooling information across subjects, we estimate spatial correlation functions that capture within-type and cross-type dependencies, enabling interpretable inference about disease-specific cellular organization. We validate our method using simulations, demonstrating robustness to latent factor specification and spatial resolution. We apply our approach to two multiplexed imaging datasets: pancreatic cancer and colorectal cancer, revealing distinct spatial organization patterns across disease subtypes and highlighting tumor-immune interactions that differentiate immune-permissive and immune-exclusive microenvironments. These findings provide insight into mechanisms of immune evasion and may inform novel therapeutic strategies. Our approach offers a principled framework for modeling spatial dependencies in multi-subject data, with broader applicability to spatially resolved omics and imaging studies. An R package, available online, implements our methods.
\end{abstract}

\noindent%
{\it Keywords:}  Hierarchical Bayesian modeling, cell-cell interactions, tumor microenvironment, spatial correlation functions, log-Gaussian Cox processes.
\vfill

\newpage
\spacingset{1.9} 
\section{Introduction}\label{sec: Intro}
The tumor microenvironment (TME) is a highly complex and heterogeneous ecosystem composed of diverse cell types, vascular structures, and extracellular components \citep{quail_microenvironmental_2013,najafi_tumor_2019,arneth_tumor_2019,labani-motlagh_tumor_2020}. Spatial organization within the TME plays a critical role in oncogenesis, tumor progression, and treatment response \citep{ni_role_2021,wang_role_2023b,giraldo_clinical_2019}. Cells interact within spatial niches, forming disease-specific co-location patterns that influence tumor-immune interactions \citep{yuan_spatial_2016,tsujikawa_prognostic_2020,de_visser_evolving_2023}. Multiplexed imaging technologies, such as multiplexed immunofluorescence (mIF), have become essential tools for studying spatial organization in the TME. These techniques provide high-resolution spatially resolved single-cell data, enabling researchers to quantify how different cell types interact within diseased tissue \citep{sheng_multiplex_2023}. 

To fully understand the spatial features of the TME, we must infer both the disease-specific spatial dependence of individual cell types as well as their cross-dependence with other cell types. Multivariate geostatistical models achieve this goal via the estimation of parametric marginal and cross-correlation functions, which describe how correlation decays with distance \citep{banerjee_hierarchical_2014, genton_cross-covariance_2015}. Slow decay indicates strong spatial dependence, while rapid decay suggests weak dependence. Marginal correlation functions capture associations within the same cell type, while cross-correlation functions capture associations between different types. Together, these functions establish all pairwise dependencies, making multivariate geostatistical models a comprehensive and highly interpretable framework for characterizing spatial relationships among cell types. 


However, despite the availability of multi-subject multiplexed imaging datasets, the statistical analysis of multi-subject spatial data presents several key challenges. 
First, standard multivariate geostatistical and point process models are computationally challenging, and thus often limited to single-patient studies, leading to an inability to generalize spatial patterns across cohorts. 


Existing methods that integrate multi-subject imaging data typically ignore spatial dependencies or rely on ad hoc spatial analyses. For instance, spatially-aware models for multi-subject brain imaging data have been explored \citep{xu_modeling_2009,sanyal_bayesian_2012}, but these are univariate methods that lack the ability to simultaneously model the spatial characteristics of multiple cell types. 
Some studies use factor models to pool multi-subject multivariate data \citep{de_vito_multi-study_2019,vito_bayesian_2021,wang_multiple_2024,lock_bidimensional_2022}, but these approaches lack spatial correlation modeling, preventing explicit characterization of cell-cell co-location patterns. Finally, machine learning or artificial intelligence pipelines used in spatial tissue analysis often function as black boxes that lead to opaque interpretation, prevent reliable uncertainty quantification, or lack the ability to pool information from multiple subjects \citep{hu_spagcn_2021,mckinley2022miriam,wu_graph_2022}, while others are software tools developed for image processing and visualization and thus lack a statistical model for inference \citep{schapiro2022mcmicro, warchol_visinity_2023}.


To address these challenges, we introduce 
a novel Bayesian hierarchical model for multi-subject multiplexed imaging data. We model each TME image as a log-Gaussian Cox process (LGCP, \citealt{diggle_spatial_2013}). In this model, an inhomogeneous Poisson process governs the intensity of each cell type. The process intensity is spatially-varying by means of a latent multivariate Gaussian process (GP) which captures both spatial and cross-type dependence.
A key feature of our approach is its multi-subject pooling strategy. In our formulation, the spatial parameters are shared across all subjects, allowing efficient information borrowing while maintaining flexibility for localized heterogeneity. The output of our model is an estimate of all marginal and cross-correlation functions among all cell types, leading to easily interpretable inference about the spatial organization of the TME. This makes our method particularly useful for learning about disease-specific spatial organization of the TME in cohorts with limited per-patient sample sizes.

We analyze two multiplexed imaging datasets to investigate spatial dependencies in the tumor microenvironment. The first dataset, from the University of Michigan’s Pancreatic Cancer Center (UM-PCC), consists of multiplexed images from pancreatic cancer patients \citep{krishnan_gawrdenmap_2022,krishnan_proximogrammulti-omics_2024}. The second dataset is a publicly available colorectal cancer (CRC) dataset, which includes 35 patients with distinct histopathological subtypes \citep{schurch_coordinated_2020}.  

Our approach provides a principled statistical framework for detecting disease-specific spatial patterns, uncovering tumor-immune interactions, and assessing their implications for cancer progression and treatment response. After introducing the datasets and our preprocessing workflow in Section \ref{sc:data}, we present our method in Section \ref{sc:methods} and demonstrate its ability to estimate cross-correlation functions via simulations in Section \ref{sc:simulations}. We then apply our method to the UM-PCC dataset in Section \ref{sc:panc} and to the CRC dataset in Section \ref{sc:crc}. Finally, we summarize our contributions in the discussion section. The Supplement includes additional methodological details, simulation results, sensitivity analyses, guidelines for model parameter selection, and further output from our UM-PCC and CRC analyses. An \texttt{R} software package for implementing our methods is available at \rpack, with code to reproduce our analyses at \codereproduce.

\section{Many-subject multiplexed imaging datasets}\label{sc:data}

\begin{sidewaysfigure}
    \centering
    \includegraphics[width=\textwidth]{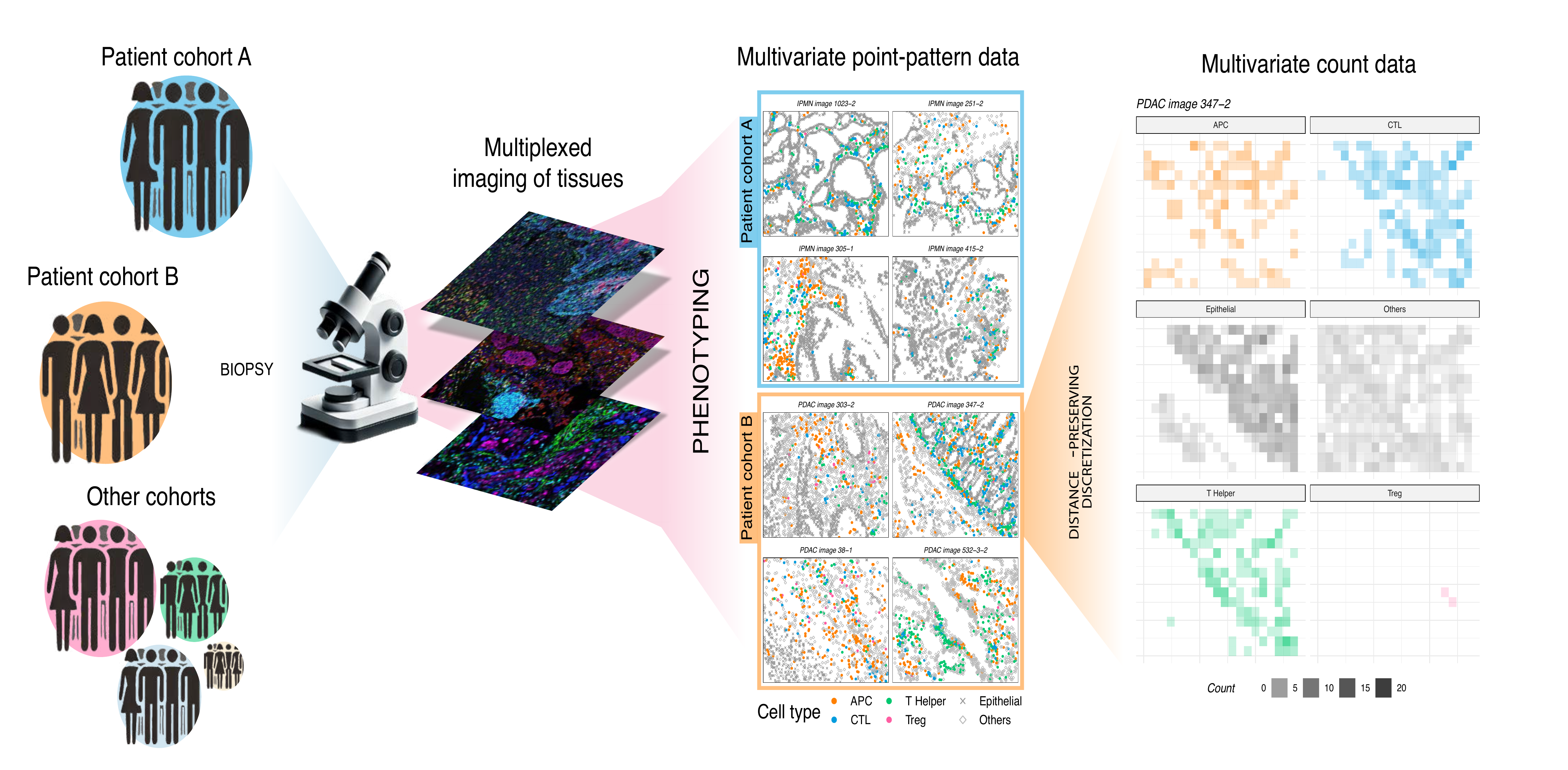}
    \caption{Overview of the imaging and statistical analysis workflow. Biopsy samples are first collected from patients across different cohorts. These biopsies are then processed and imaged using multiplex imaging techniques, allowing for high-dimensional spatial characterization of the tumor microenvironment. Following imaging, cell phenotyping is performed to classify cells based on marker expression, generating spatial point patterns for each sample. These point patterns are subsequently converted into multivariate binned count data, which serves as the structured input for the statistical method presented in this study.}
    \label{fig:workflow}
\end{sidewaysfigure}

Multiplexed imaging datasets from multiple subjects enable the spatial characterization of tissue organization at single-cell resolution. Despite differences in disease context, these datasets share a common structure: each image consists of spatially localized cells annotated with phenotypic labels derived from standardized staining and imaging pipelines. The primary challenge in analyzing such datasets lies in modeling spatial dependencies across subjects while accounting for inter-patient variability in cell composition and tissue architecture. In Figure~\ref{fig:workflow}, we show the workflow for this analysis, beginning with the collection of biopsy samples from different patient cohorts, followed by multiplex imaging and cell phenotyping. The resulting spatial point patterns are then transformed into multivariate binned count data, which serves as the structured input for our method.

\subsection{Pancreatic cancer dataset}
Images from the pancreatic cancer dataset were obtained from patients who had presented at the University of Michigan Pancreatic Cancer Clinic and subsequently underwent surgical resection for various pancreatic pathologies. The study population consisted of patients diagnosed with six distinct pancreatic diseases, encompassing both malignant and non-malignant conditions. Our primary focus was on two specific patient groups: intraductal papillary mucinous neoplasm (IPMN, 70 patients) and pancreatic ductal adenocarcinoma (PDAC, 71 patients).

\begin{figure}[!ht]
    \centering
    \includegraphics[width=\textwidth]{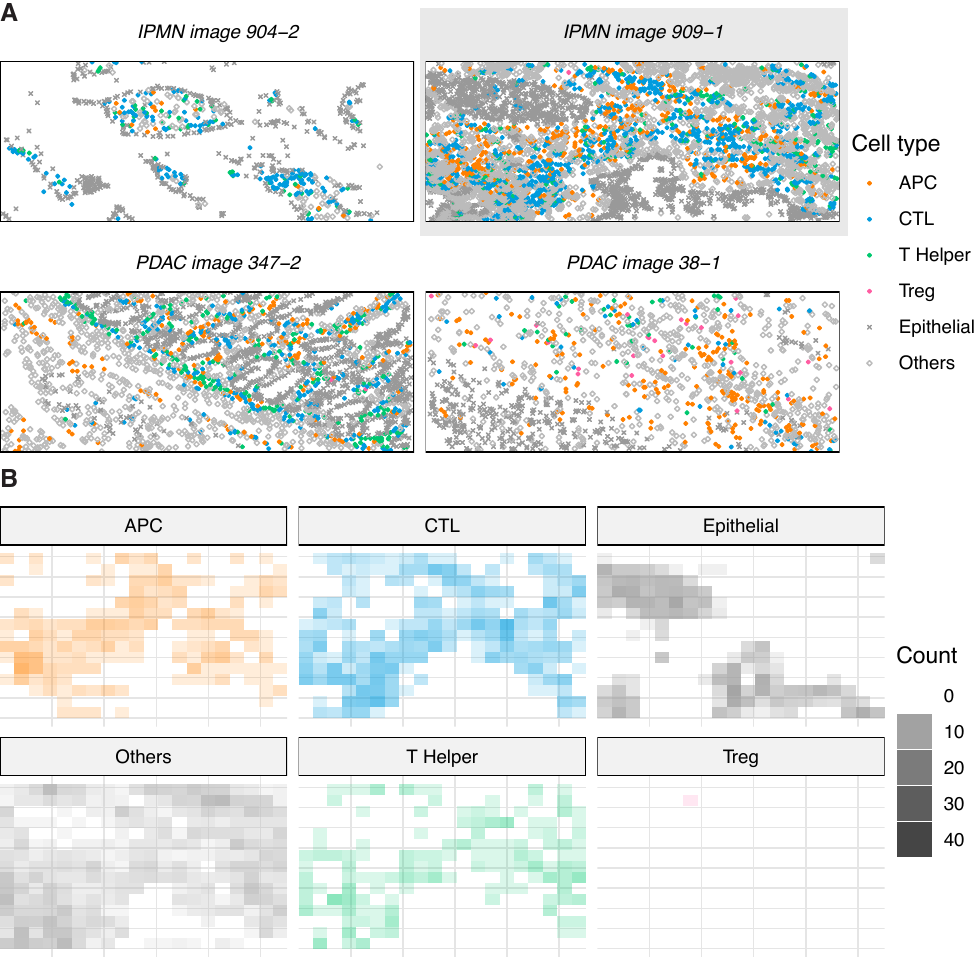}
    \caption{Point patterns derived from selected images in the pancreatic cancer dataset, including both IPMN and PDAC cases. A) Two images from each cancer type. B) IPMN image 909-1 is transformed into the binned count representation.}
    \label{fig:panc_pat_count}
\end{figure}

Tissue images were obtained via multiplexed immunofluorescence (mIF) staining, which was performed following protocols outlined in \cite{lazarus_spatial_2018}. Processing and subsequent phenotyping were carried out using the methods described by \cite{krishnan_gawrdenmap_2022}.
 Phenotypic classification was performed using user-defined thresholds and standardized with inForm Tissue Analysis software \citep{akoya_inform_2021} across the six original disease groups. Each cell was classified into one of six phenotypes: antigen-presenting cell (APC), cytotoxic T lymphocyte (CTL), epithelial (Epithelial), helper T cell (T Helper), regulatory T cell (Treg), or others. Figure~\ref{fig:panc_pat_count} illustrates representative point patterns from the pancreatic cancer dataset, highlighting spatial distributions of these phenotypes in selected IPMN and PDAC images. The bottom row further demonstrates the transformation of an IPMN point pattern (sample 909-1) into a binned count representation, which aggregates local cell densities to serve as input for subsequent statistical analysis by our method.

\subsection{Colorectal cancer dataset}

The second dataset we consider consists of multiplexed images of tumor tissue samples sourced from a publicly available colorectal cancer repository \citep{schurch_coordinated_2020}. These images depict various cell types within the tumor microenvironment, including malignant epithelial cells, various immune subtypes and other stromal cells. 
The PhenoCycler imaging technology~\citep{goltsev_deep_2018,black_codex_2021} was used to generate this dataset, which consists of multiplexed imaging data from 35 CRC patients, each contributing four biopsies. A single high-dimensional image was acquired per biopsy, capturing the expression of 56 distinct proteins across tumor tissue sections. Patients were categorized based on tumor histopathology into two groups: those with visible tertiary lymphoid structures (TLSs) at the invasive front (CLR, Crohn’s-like reaction) and those with diffuse inflammatory infiltration (DII) without detectable TLSs. CLR patients generally exhibit an immune-hot, well-mixed tumor microenvironment (TME) and a more favorable prognosis, whereas DII patients have an immune-exclusive, cold TME~\citep{schurch_coordinated_2020}.

\begin{figure}[!ht]
    \centering
    \includegraphics[width=\textwidth]{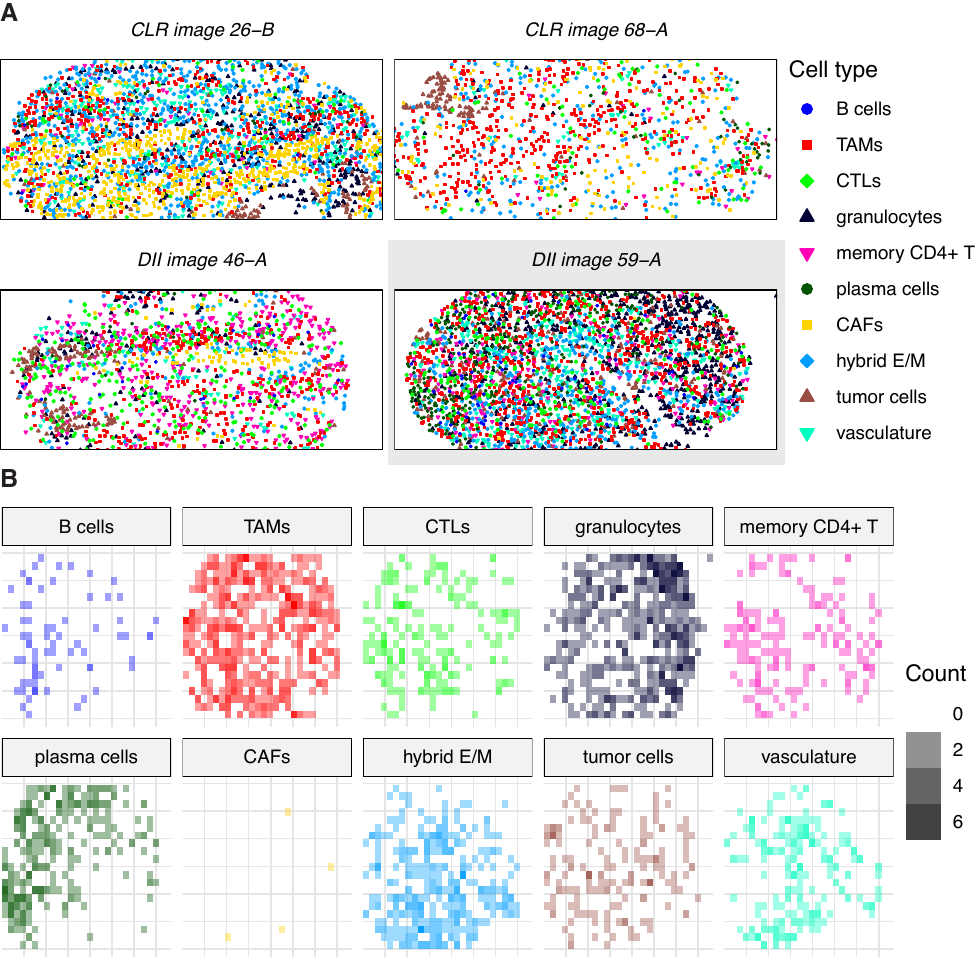}
        \caption{\label{fig:crc_pat_count}
        Point patterns derived from selected images in the colorectal cancer dataset, including both CLR and DII cases. A) Two images from each cancer type. B) DII image 59-A is transformed into the binned count representation.}
\end{figure}

The original study annotated 16 cell types across nearly 200,000 cells, but we focus on the 10 most common. We also relabeled “stroma” cells as “hybrid E/M” due to co-expression of Vimentin and cytokeratin, indicating both epithelial and mesenchymal traits~\citep{kuburich_cancer_2024}. Similarly, “smooth muscle” cells were reclassified as cancer-associated fibroblasts (CAFs) based on $\alpha$-SMA and Vimentin expression~\citep{cao_cancer-associated_2025}. We excluded the six least frequent cell types from the CRC analysis because their rarity limits their contribution to spatial modeling. Since these cell types provide minimal spatial information, we focused on the more prevalent populations. In future work, we may extend the model to incorporate non-spatial components to account for these rare cell types. Figure~\ref{fig:crc_pat_count} presents representative point patterns from the colorectal cancer dataset, illustrating spatial distributions in selected CLR and DII cases, with the bottom row showing the binned count representation for sample 59-A.

\section{Multi subject Bayesian latent spatial factor model}\label{sc:methods}
Let $\calD_i \subset \Re^2$ represent the $i$th multiplexed image in the data, with $i=1,\dots, N$. Moving forward, we assume that each subject is associated to a single image to simplify exposition and without loss of generality. Then, let $\calY_{ij} = \{ Y_{ij}(\bl) : \bl \in \calD \}$ be a subject-specific inhomogenous Poisson process related to the point-pattern of cells of type $j$, where $j=1, \dots, q$. We then introduce $\calW = \{ \bw(\bl): \bl \in \calD \}$ as a zero-mean multivariate GP and denote as $w_{ij}(\bl)$ the realization at $\bl$ of its $j$th margin for subject $i$. $\calW$ is completely defined by the user-specified cross-covariance matrix function indexed by parameter vector $\btheta$ and defined as $\Cov(\cdot, \cdot; \btheta): \calD\times\calD \rightarrow \boldsymbol{M}_{q\times q}$, where $\boldsymbol{M}_{q\times q}$ is the set of $q \times q$ real-valued matrices, and such that $\Cov(\bl,\bl'; \btheta) = \Cov(\bl',\bl; \btheta)^{\top}$ and $\sum_{i=1}^n\sum_{j=1}^n \bx_i^{\top}\Cov(\bl_i,\bl_j; \btheta)\bx_j > 0$ for any integer $n$, any finite set $\{\bl_1,\bl_2,\ldots,\bl_n\}$, and for all $\bx_i,\bx_j \in \Re^q\setminus \{\bzero\}$ \citep{genton_cross-covariance_2015}. Therefore, by construction, the $(r, s)$th entry of $\Cov(\bl,\bl')$ evaluates to $C_{rs}(\bl, \bl';\btheta)=\text{cov}(w_r(\bl),w_s(\bl'))$. Under an assumption of stationarity, we have $\Cov(\bl, \bl'; \btheta) = \Cov(\bh; \btheta)$, where $\bh = \| \bl-\bl' \|$. 
We then complete the LGCP by assuming that, conditional on $\calW$, the intensity of $\calY_{ij}$ at location $\bl$ is $\Lambda_{ij}(\bl) = \exp \{ \bx_i^\top(\bl)\bbeta_j + w_{ij}(\bl) \}$. We choose a spatial factor model as our cross-covariance matrix function:  $C_{rs}(\bl, \bl') = \sum_{j=1}^k \lambda_{rj}\lambda_{sj} \rho(\bl, \bl'; \varphi_j)$, where $\lambda_{rj}$ is the $(r,j)$ element of the $q \times k$ matrix $\bA$ of factor loadings, $\rho(\bl, \bl'; \varphi_j) = \exp\{ -\varphi_j \|\bl-\bl' \| \}$, and therefore $\btheta = \{ \bA, \varphi_1, \dots, \varphi_k \}$.

Our model specification has two main features: first, our conditional independence assumption implies that $\calY_{ij}(\bl)$ incur in spatial and cross-cell-type dependence by means of marginalization of $\calW$. Our primary inferential targets are the marginal correlation functions $\rho_{rr}(\bh; \btheta) = \frac{C_{rr}(\bh;\btheta)}{C_{rr}(\bzero; \btheta)}$ and the cross-correlation functions $\rho_{rs}(\bh; \btheta) = \frac{C_{rs}(\bh; \btheta)}{\sqrt{C_{rr}(\bzero; \btheta)} \sqrt{C_{ss}(\bzero; \btheta)}}$, where $r\neq s$. These functions (which total $q(q+1)/2$ in number) give an easily interpretable characterization of the joint spatial pattern of the log-intensity process among all cell types. 
The second main feature of our model is that we assume $\btheta$ is common across \textit{all} subjects in the same group. Because we group subjects by disease and $\btheta$ determines the shape of all correlation functions $\rho_{rs}(\cdot; \btheta)$ for $r,s=1, \dots, q$, our model pools information from all subjects to characterize the disease-specific joint point pattern of cell types.

Because computations involving LGCPs are highly complex even in univariate single-subject settings, we follow \cite{diggle_spatial_2013} and discretize the image using a computational grid of dimension $n_{x}\times n_{y}$, for a total of $n_{px}$ pixels. Suppose $\bl$ and $\bl'$ are two locations in any of the images in the data. The preprocessing pipeline we described in Section \ref{sec:preprocessing} ensures that $\bh=\| \bl - \bl' \|$ is the same for all images and therefore ensure that each image correctly contributes to learning the common cross-covariance matrix function $\Cov(\cdot; \btheta)$. 
We then refer to $y_{ij}(\bl)$ as the number of cells of type $j$ at location $\bl$ in the $i$th image. The LGCP model on the discretized images translates to
\begin{equation}\label{eq:multi_subject_model}
\begin{aligned}
y_{ij}(\bl) \mid \lambda_{ij}(\bl) \sim \text{Poisson}(\lambda_{ij}(\bl))&, \qquad \blambda_i(\bl) = \begin{bmatrix} \lambda_{i1}(\bl) \cdots \lambda_{iq}(\bl) \end{bmatrix}^\top\\
\log\blambda_i(\bl) = \balpha_i + \bX_i(\bl)\bB + \bw_i(\bl)&, \qquad \bw_i(\bl)=\bA \bv_i(\bl),\\
\bv_i(\bl) = \begin{bmatrix} v_{i,1}(\bl), \dots, v_{i,k}(\bl) \end{bmatrix},&^\top \qquad v_{i,j}(\cdot) \sim GP(\bzero, \rho(\cdot,\cdot; \varphi_j)),
\end{aligned}
\end{equation}
where we assign zero-mean Gaussian priors to $\balpha_i$, each column of $\bB$, and each row of $\bA$. Finally, we assign the prior $\varphi_j \sim U(0.1, 10), j=1,\dots,k$ to allow a wide range of possible spatial dependence.  In \eqref{eq:multi_subject_model}, $\balpha_i$ is a cell-type-specific offset for the overall intensity, whereas the term $\bw_i(\bl)=\bA \bv_i(\bl)$ represents a latent spatial factor model defined above. Because the latent spatial effects $\bw_i(\bl)$ are subject specific, our model allows variability in the observed images without requiring that they align--indeed, that would be impossible because each image corresponds to a different tissue. Nevertheless, our model borrows information across all images to estimate all $q(q+1)/2$ cross-covariance functions. 
Finally, we assume the effect of covariates $\bX_i(\bl)$ to be common across images; although relaxing this assumption is straightforward, it is beyond the scope of this article. 

GPs are notoriously computationally inefficient when the number of spatial locations is large. Although one possible solution is to discretize the raw data on a coarser computational grid (i.e., lower $n_{px}$), reducing grid size negatively impacts the ability to infer LGCP paramters \citep{diggle_spatial_2013}. Instead, we replace the GP with a meshed GP \citep[MGP, ][]{peruzzi_highly_2022}. An MGP is scalable to large datasets because it restricts spatial dependence using a user-specified sparse directed acyclic graph (DAG) which operates on a partition of the spatial domain. Intuitively, each domain subregion is independent of non-neighboring regions, given its neighbors. Scalable GPs based on sparse DAGs have become extremely popular in geostatistical settings due to their numerous desirable properties; notably, they lead to algorithms that scale linearly with the number of observed spatial locations, \citep{vecchia_estimation_1988, datta_hierarchical_2016, katzfuss_general_2021}, with additional computational savings on imaging data \citep{peruzzi_highly_2022}, and with theoretical guarantees regarding their ability to approximate unrestricted GPs \citep{zhu_radial_2024}. Therefore, any such DAG-based GP can scalably approximate an unrestricted GP, with inferential performance being indistinguishable in practice. We fit MGPs to our  datasets via Markov-chain Monte Carlo methods for multivariate count data, extending those outlined in \cite{peruzzi_spatial_2024} to the multi-subject setting. More details on the MCMC algorithm are in the Supplement.

\subsection{Data preprocessing workflow} \label{sec:preprocessing}

To ensure compatibility with our method and its software implementation, we first standardized the spatial coordinates of each image. Specifically, we first shifted each image so that its bottom-left corner aligns with the origin \((0,0)\). Next, we rescaled the coordinates to normalize image dimensions across the dataset. Let \( l_{x,i} \) and \( l_{y,i} \) represent the $x$-axis and $y$-axis lengths (in microns) of the \( i \)th image. We computed the maximum axis length across all images as $l^* = \max\{ l_{x,i}, l_{y,i} \}.$ 
Each image was then rescaled so that its spatial domain is given by $\mathcal{D}_i = \left[0, \frac{l_{x,i}}{l^*} \right] \times \left[0, \frac{l_{y,i}}{l^*} \right] \subset [0,1]^2.$
This transformation ensures that all images share a common unit of measurement while preserving relative distances within each image. This rescaling step simplifies the specification of priors for spatial range parameters while maintaining the integrity of spatial relationships.  

It is customary to fit LGCPs by first discretizing the raw point-pattern data into a high resolution pixel image where each pixel is a vector of cell-type counts \citep{diggle_spatial_2013}. Therefore, we constructed an \( n_x \times n_y \) grid over each image and counted the number of cells of each type within each grid pixel. We denote the resulting count data as  
$\by_i(\bl)$, a vector of dimension \( q \), with its \( j \)th element representing the number of cells of type \( j \) at location \( \bl \) in the \( i \)th image. Examples of the processed images after these transformations are shown in Figure~\ref{fig:panc_pat_count}B and Figure~\ref{fig:crc_pat_count}B.

\section{Simulation studies}
\label{sc:simulations}

We performed two simulation studies in which we examined the impact of hyperparameters such as the number of latent factors $k$ and the number of grid pixels $n_{px}$ on inference quality.
In particular, we studied the effect of misspecification of $k$ on the convergence of MCMC sampling, the recovery of cross-correlation curves between cell types, and the use of information criteria to appropriately choose $k$. In the second study, we consider the case where we use a coarser grid of pixels than is assumed in the generative model, again focusing on how MCMC sampling and cross-correlation curve recovery are affected. In both of these simulation studies, we quantified the degree of cross-correlation curve recovery via the median absolute deviation (MAD) between the observed and fitted cross-correlation curves.

\subsection{Recovering cross-correlations when the number of latent factors is misspecified}

\begin{figure}
\centering
\includegraphics[width=\textwidth]{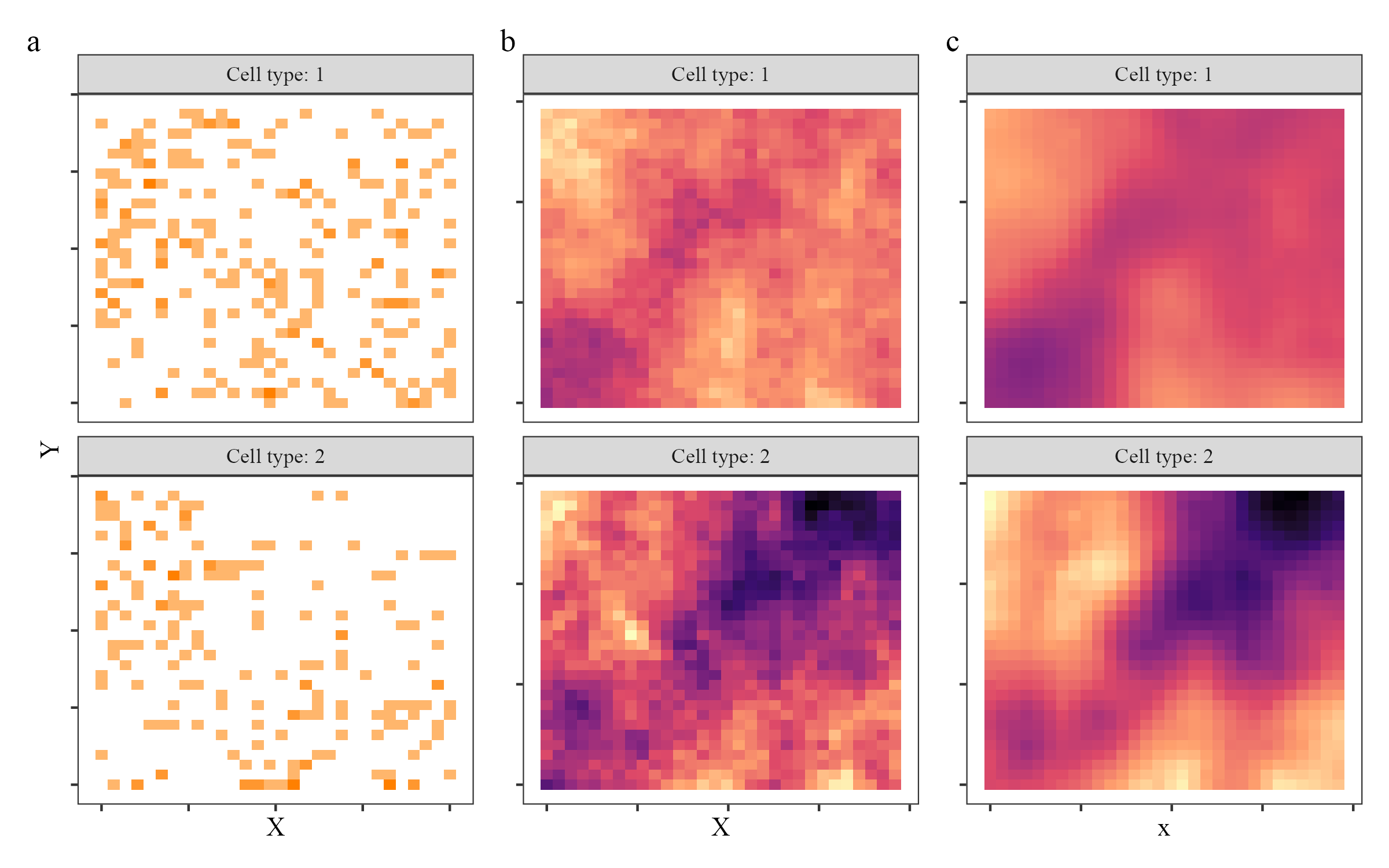}
\caption{Example simulation from study on misspecification of $k$. a) Simulated data for a single image. b) Underlying latent factors of simulated data. c) Recovered latent factors.}
\label{fig:pred_W_obs_W_k}
\vspace{1mm}
\end{figure}

In this set of simulations, we set the true, generative $k$ (what we call $k^*$) as $k^*=3$ and then examined how the recovery of parameters is affected as we fit the model with $k\in\{2,3,6\}$. For each $k$, we simulated 10 datasets of $N=40$ images each, where each image is a $30\times 30$ pixel grid of 10 cell types, with pixel coordinates for all images from the rectangular domain $[0,1]\times [0,0.75]$. We employed an exponential covariance function for the spatial latent factors $\bv_i(\bl)$, in which the spatial decay $\varphi_j$ of each factor $v_{i,j}(\bl)$ was sampled uniformly from $[1,3]$. Lastly, we used an intensity offset of $\alpha_{ij}=-2$ for each $i=1,\dots, N$ and $j=1,\dots q$. We then simulated from our model--refer to the Supplement for additional details.
After generation of ground-truth simulations, we fit our model to these simulations and sampled 5000 MCMC samples after a 5000-sample burn-in period. An example of simulated data can be seen in Figure \ref{fig:pred_W_obs_W_k}a, while an example of the underlying latent factors $\bw_i(\bl)$ can be seen in Figure \ref{fig:pred_W_obs_W_k}b.

\begin{figure}
\centering
\includegraphics[width=\textwidth]{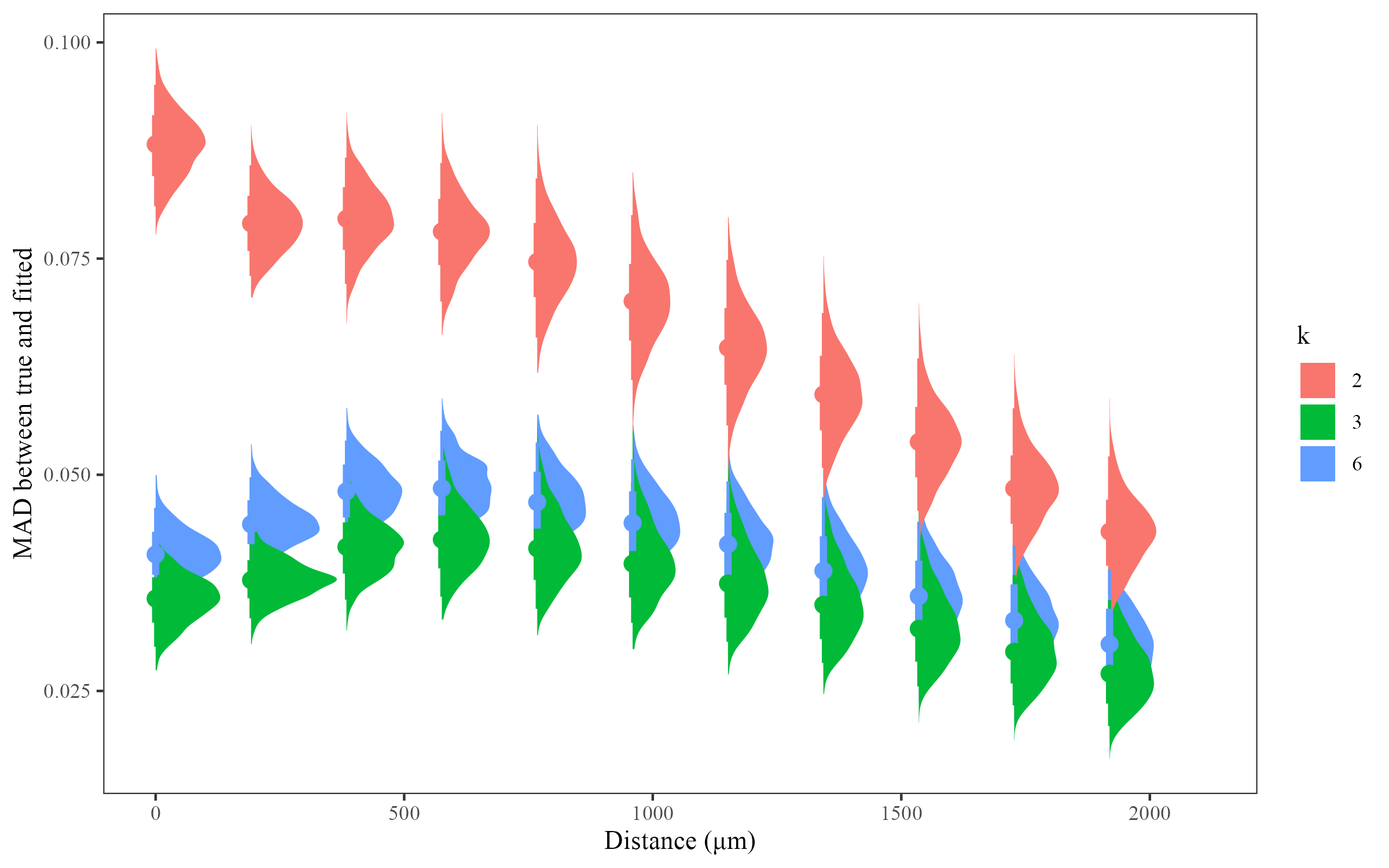}
\caption{MAD of cross-correlation curves at various distances between observed and predicted when fitting with various $k$. Here the generative model has $k^*=3$.}
\label{fig:mad_distance_k}
\vspace{1mm}
\end{figure}

Recovery of the cross-correlation curves was best when $k=k^*=3$ (Figure \ref{fig:mad_distance_k}). If $k$ was misspecified, we found that $k>k^*$ yielded cross-correlation curves that were very similar to the true curves (Figure \ref{fig:mad_distance_k}). Underspecified $k$ had the worst recovery at closer spatial distances. Convergence diagnostics, such as effective sample size (ESS) \citep{geyer_introduction_2011} and Gelman's $\hat{R}$ \citep{vehtari_rank-normalization_2021}, however, showed that smaller $k$ demonstrated better MCMC convergence, even when $k < k^*$. In general, MCMC convergence of cross-correlation estimates was the best for zero-distance cross-correlations. Lastly, we applied the Watanabe-Akaike information criterion (WAIC) \citep{watanabe_asymptotic_2010} to models fit under various $k$ and found, again, that underspecified $k$ leads to the worst predicted out-of-sample predictive performance, while fitting with over-specified $k$ has a WAIC that is very similar to that of models fit with $k^*$. Based on these results, in general we recommend that users perform model fitting at various $k$ and then use the model at a value of $k$ for which the WAIC is no longer improving. Doing so will ensure the best out-of-sample predictive performance, as well as high-fidelity parameter recovery. The Supplement includes additional details and Figures related to our simulation study.

\subsection{The effect of grid size on cross-correlation recovery}

\begin{figure}
\centering
\includegraphics[width=\textwidth]{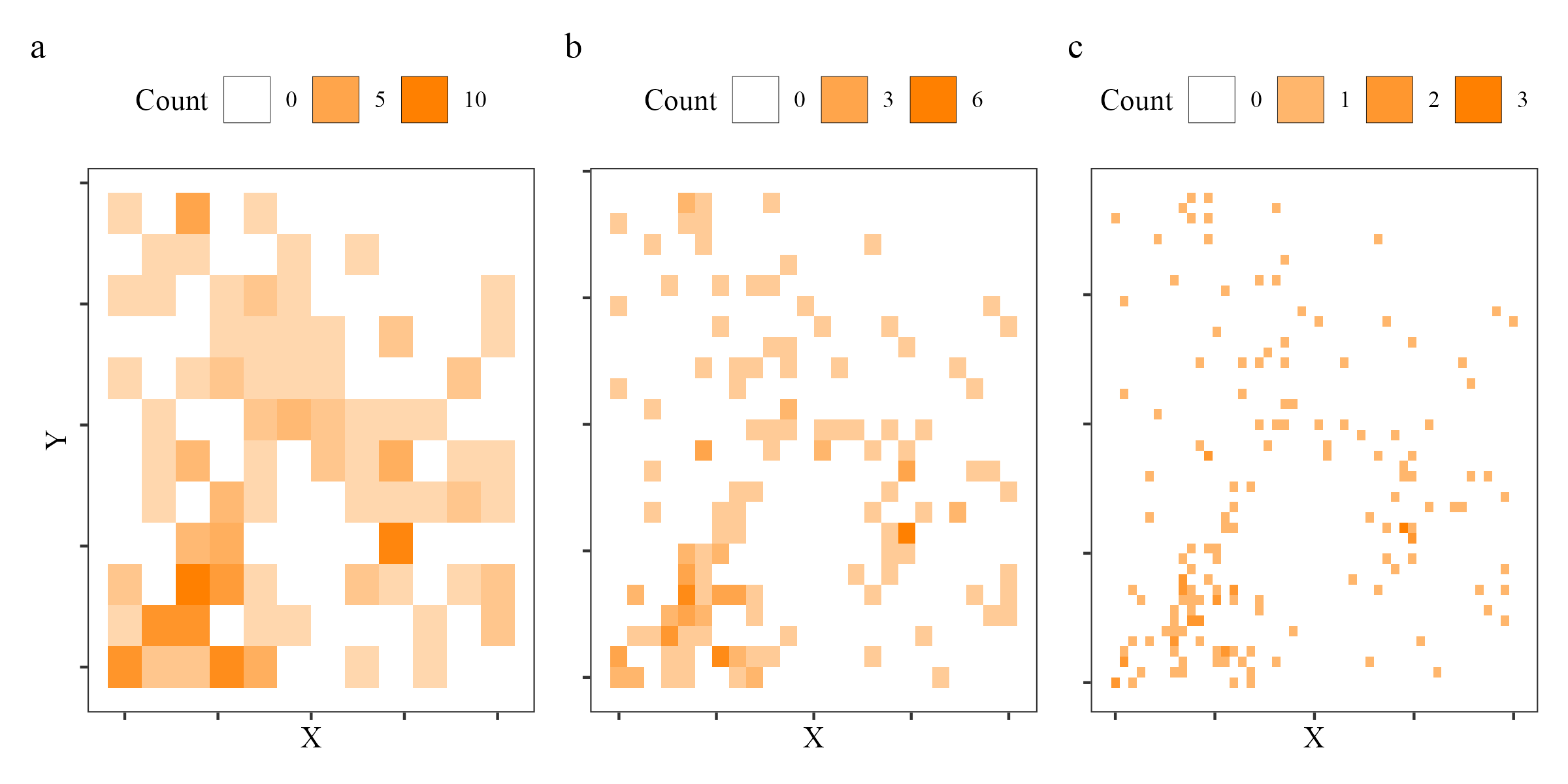}
\caption{ Simulated data for a single image with a) $n_x=12$, b) $n_x=24$ and c) $n_x=48$.}
\label{fig:Y_list_diff_res}
\vspace{1mm}
\end{figure}

In the next set of simulations, we set  $k^*=4$ and then examined how the recovery of parameters is affected as we fit the model with number of pixels on each dimension $n_{x}\in{12,24,48}$. We also set $n_y=n_x$, so that each image is $n_x\times n_x$. For each $n_{x}$, we simulated 10 datasets of $N=40$ images each, where each image was a $48\times 48$ pixel grid of 10 cell types, with pixel coordinates for all images from the rectangular domain $[0,1]\times [0,0.75]$. We then coarsened each pixel grid to an $n_{x}\times n_{x}$ grid by aggregating neighboring pixels together. We used an intensity offset of $\alpha_{ij}=-3$ for each $i=1,\dots, N$ and $j=1,\dots q$. We kept the rest of the parameters the same as from the last simulation--specifically, we used spatial latent factors $\bv_i(\bl)$ with exponential covariance and sampled each spatial decay parameter $\varphi_j$ from $U(1,3)$. We then simulated from our model--additional details are in the Supplement. After generation of ground-truth simulations, we fit our model to these simulations and sampled 1000 MCMC samples thinned every 5 draws after a 5000-draw burn-in period. An example of simulated data at various resolutions can be seen in Figure \ref{fig:Y_list_diff_res}.

\begin{figure}
\centering
\includegraphics[width=\textwidth]{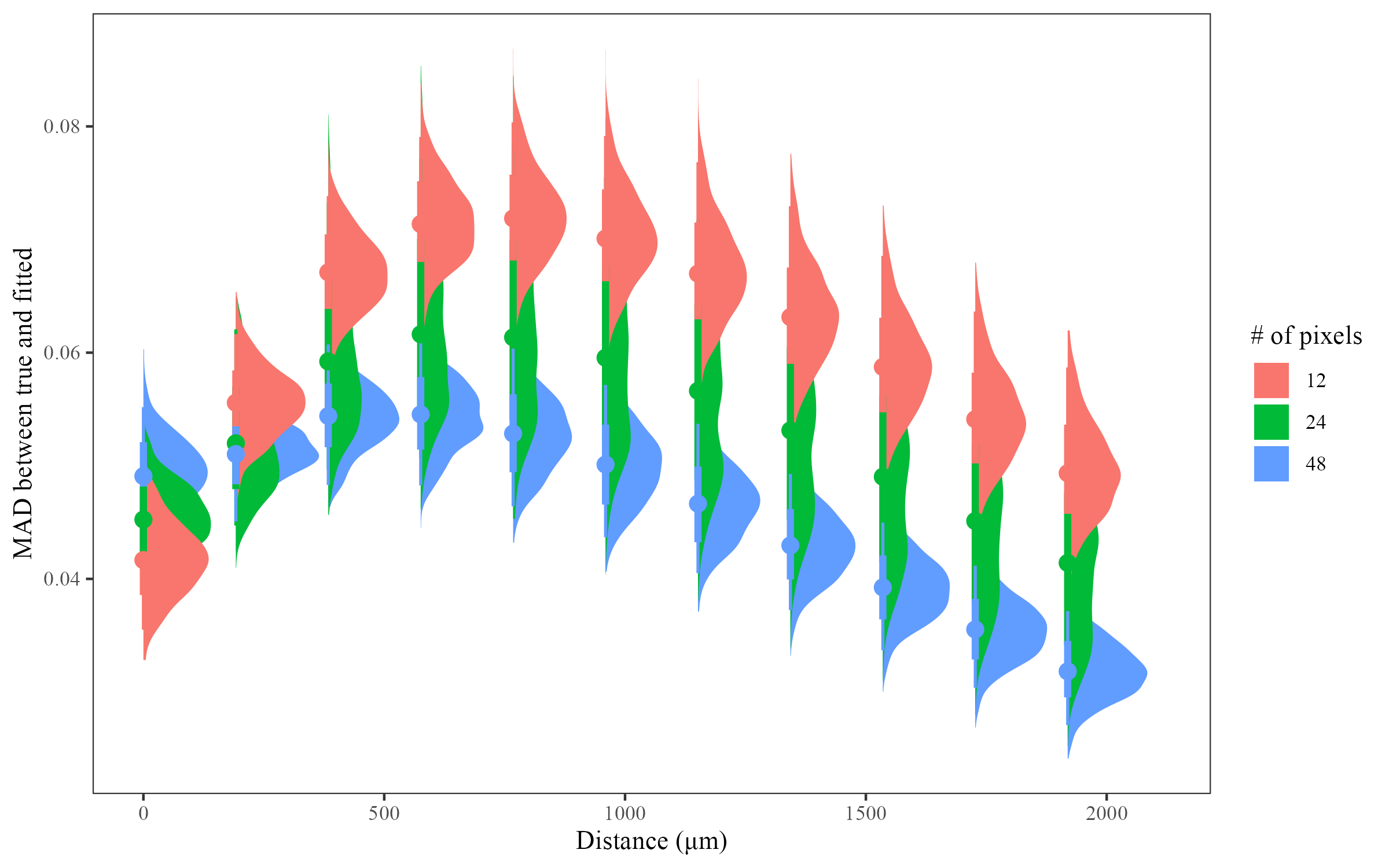}
\caption{MAD of cross-correlation curves at various distances between observed and predicted when fitting with various numbers of pixels.}
\label{fig:mad_distance_sz}
\vspace{1mm}
\end{figure}

We found that the MAD between observed and fitted was much closer between simulations with different $n_{x}$ than in the previous simulation suite, and seems to follow a similar trend over spatial distance: increasing up to about 1000 $\mu m$ and then decreasing (Figure \ref{fig:mad_distance_sz}). Models fit with the largest number of pixels $n_{x}=48$ had nominally the best cross-correlation curve recovery. In the Supplement, however, we show that the convergence diagnostics for models fit on data with higher numbers of pixels were significantly worse, and that cross-correlation curves fit on different $n_{x}$ seem to be visually indistinguishable from one another. Finally, the use of an information criterion was not useful to us in this situation, as the datasets we would be comparing are of different sizes, but we can recommend that users wanting to determine the appropriate number of pixels should run a sensitivity analysis at various pixel count and determine the coarsest number of pixels that they can use without affecting the cross-correlation curve recovery too much.

\section{Application to pancreatic cancer data}\label{sc:panc}

We applied our method to the full set of 143 images in the PDAC cohort and 89 images in the IPMN cohort, focusing on key cell types such as regulatory T cells (Tregs), helper T cells, epithelial (tumor) cells, cytotoxic T lymphocytes (CTLs), antigen-presenting cells (APCs), and others. Each image was scaled and divided into a grid of pixels, each of which was approximately $70\times 70\mu m$, following our data preprocessing workflow (see Figure \ref{fig:panc_pat_count}B for an example of a preprocessed image). After preprocessing, we fit our model separately to the PDAC and IPMN cohorts, testing $k\in\{2,3,4,5\}$ factors. We run MCMC for 30,000 iterations, discarding the first 20,000 as burn-in and thinning 10:1 the remaining ones, retaining a total of 1,000 MCMC samples.

WAIC values for both patient groups were lowest for $k=5$ (Table \ref{tab:waic_panc}). However, results from the $\hat{R}$ diagnostic were mixed - no single model had universally better convergence of cross-correlation curves for all cell type pairs. In both groups, the $\hat{R}$ of the cross-correlation curves for most cell type pairs at $k=5$ were quite good, indicating acceptable MCMC convergence. Thus, we chose $k=5$ for both models. The Supplement includes additional details on MCMC diagnostics. 

\begin{table}[!h]
\centering
\caption{\label{tab:waic_panc}The WAIC of the fitted model for various $k$, for both patient groups in the pancreatic cancer dataset.}
\centering
\resizebox{\ifdim\width>\linewidth\linewidth\else\width\fi}{!}{
\begin{tabular}[t]{>{}ccc}
\toprule
\multicolumn{1}{c}{ } & \multicolumn{2}{c}{WAIC Scores} \\
\cmidrule(l{3pt}r{3pt}){2-3}
\textbf{k} & \textbf{WAIC in PDAC group} & \textbf{WAIC in IPMN group}\\
\midrule
\textbf{\cellcolor{gray!10}{2}} & \cellcolor{gray!10}{3197} & \cellcolor{gray!10}{2933}\\
\textbf{3} & 3090 & 2941\\
\textbf{\cellcolor{gray!10}{4}} & \cellcolor{gray!10}{2958} & \cellcolor{gray!10}{2894}\\
\textbf{5} & 2940 & 2843\\
\bottomrule
\end{tabular}}
\end{table}

\subsection{Spatial Organization of the Pancreatic Cancer TME Across Patient Groups}

\begin{sidewaysfigure}
\centering
\includegraphics[width=\textwidth]{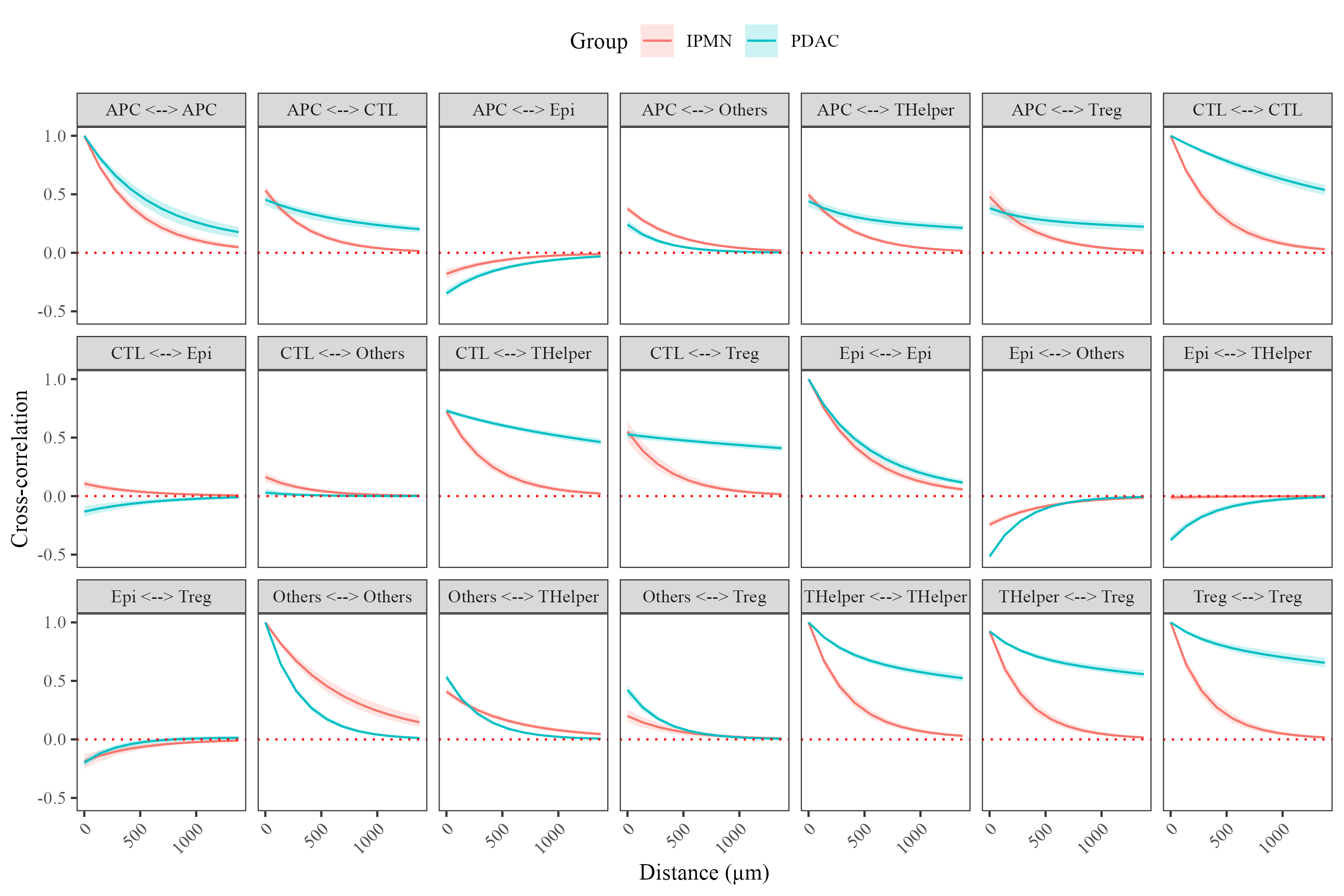}
\caption{The estimated spatial cross-correlations between all pairs of cell-type combinations from patients in the IPMN group and the PDAC group. The colored line in each plot indicates the posterior expectation, while the shaded region represents the pointwise 95\% credible interval.}
\label{fig:xcor_panc}
\vspace{1mm}
\end{sidewaysfigure}

The spatial cross-correlation analysis reveals distinct cellular organization patterns between PDAC and IPMN tumors, highlighting key differences in immune cell spatial distribution (Figure~\ref{fig:xcor_panc}). In general, PDAC tumors exhibit stronger spatial associations among key immune cell populations, whereas IPMN tumors display weaker correlations, indicating a less structured immune landscape.

\paragraph{Regulatory T Cell Expansion and Clustering in PDAC}

A striking finding is the significantly higher spatial correlation of regulatory T cells (Tregs) in PDAC tumors compared to IPMN tumors. While Tregs show strong self-association in both groups at close distances, their correlation persists across greater spatial ranges in PDAC, suggesting extensive clustering. This spatial pattern may indicate an immunosuppressive niche within PDAC tumors, where Tregs accumulate to dampen anti-tumor immune responses. Such clustering has been implicated in immune evasion and resistance to immunotherapy in pancreatic cancer~\citep{mota_reyes_regulatory_2022}.

\paragraph{Stronger T Cell Interactions in PDAC Tumors}

Cytotoxic T lymphocytes (CTLs) and helper T cells (THelper) also display stronger spatial correlations in PDAC compared to IPMN. This pattern suggests that T cells in PDAC tumors are more spatially organized and less dispersed, potentially reflecting their recruitment into structured immune environments. The increased correlation between CTLs and THelpers further suggests coordinated immune activation in PDAC tumors, albeit within a potentially immunosuppressive microenvironment. 

Additionally, PDAC tumors exhibit a higher correlation between CTLs and Tregs at moderate to far distances, reinforcing the hypothesis that PDAC fosters a structured but immune-suppressive T cell landscape, where regulatory mechanisms counteract effector responses~\citep{goulart_t_2021}.

\paragraph{Divergent Antigen-Presenting Cell (APC) Distribution Between Groups}

Antigen-presenting cells (APCs) show markedly different spatial correlations with T cells between PDAC and IPMN tumors. In PDAC, APCs exhibit stronger associations with CTLs, THelpers, and Tregs, suggesting that these tumors maintain closer interactions between antigen-presenting and lymphoid cells. This organization could support both immunosuppressive and immune-activating roles, depending on the APC phenotype~\citep{poh_tumor-associated_2021}. The weaker APC-T cell associations in IPMN tumors indicate a less structured immune landscape, which may reflect differences in tumor immunogenicity or microenvironmental constraints on APC activity.

\paragraph{Conclusion}

Overall, PDAC tumors exhibit a more structured and spatially organized immune microenvironment compared to IPMN tumors. The increased clustering of Tregs, the stronger spatial correlations among T cells, and the heightened APC interactions in PDAC tumors suggest an immune landscape that is both more engaged and more tightly regulated. These findings align with previous observations that PDAC tumors often develop robust immunosuppressive mechanisms to evade immune-mediated tumor clearance. In contrast, the weaker immune correlations in IPMN tumors suggest a less compartmentalized and potentially less immunosuppressive immune landscape, which may have implications for differential treatment strategies targeting the immune microenvironment in pancreatic cancer. Interestingly, however, the spatial relationship between epithelial cells and the other cell types seemed generally much more similar - indicating that the spatial differences between these diseases were largely due to the the distributions of immune cells rather than those of precancerous and cancerous cells.

\section{Application to CRC data}\label{sc:crc}

We removed four images with extensive contiguous missing data (37\% to 53\%), as the software implementation of our method relies on dividing images into blocks with sufficient non-missing data. We identified all cell types across both patient groups that had a median count in each image greater than 10 and dropped the other cell types from the data. Lastly, we applied our preprocessing workflow to each image by converting each image to a pixel grid with pixels of size $70\times 70 \mu m$, as described previously (Figure \ref{fig:crc_pat_count}B). After data preprocessing, we fit our model separately on the images from each cohort with $k\in\{2,4,6,8,10\}$ for a total of 30,000 MCMC samples, of which 20,000 discarded as burn-in and 1000 retained after 10:1 thinning of the remainder chain.

We found that the models with the lowest WAIC were those with higher values of $k$ (Table \ref{tab:waic_crc}). However, such models fit with these higher values of $k$ also demonstrated more problems with MCMC convergence, with $k=4$ seeming to demonstrate the best MCMC convergence (in terms of $\hat{R}$). 
The Supplement includes additional details on MCMC fitting, with figures indicating that cross-correlation curves estimated with $k=2$ are signficantly different than the others, while curves estimated with higher values of $k$ look much more similar. We decided to compromise on these three aspects of model selection and choose $k=6$ for our final model.

\begin{table}
\centering
\caption{\label{tab:waic_crc}The WAIC of the fitted model for various $k$, for both patient groups in the CRC dataset.}
\centering
\resizebox{\ifdim\width>\linewidth\linewidth\else\width\fi}{!}{
\begin{tabular}[t]{>{}ccc}
\toprule
\multicolumn{1}{c}{ } & \multicolumn{2}{c}{WAIC Scores} \\
\cmidrule(l{3pt}r{3pt}){2-3}
\textbf{k} & \textbf{WAIC in CLR group} & \textbf{WAIC in DII group}\\
\midrule
\textbf{\cellcolor{gray!10}{2}} & \cellcolor{gray!10}{3121} & \cellcolor{gray!10}{5945}\\
\textbf{4} & 2913 & 5799\\
\textbf{\cellcolor{gray!10}{6}} & \cellcolor{gray!10}{2646} & \cellcolor{gray!10}{5565}\\
\textbf{8} & 2564 & 5471\\
\textbf{\cellcolor{gray!10}{10}} & \cellcolor{gray!10}{2571} & \cellcolor{gray!10}{5449}\\
\bottomrule
\end{tabular}}
\end{table}

\subsection{Spatial Organization of the CRC TME Across Patient Groups}

\begin{sidewaysfigure}
\centering
\includegraphics[width=\textwidth]{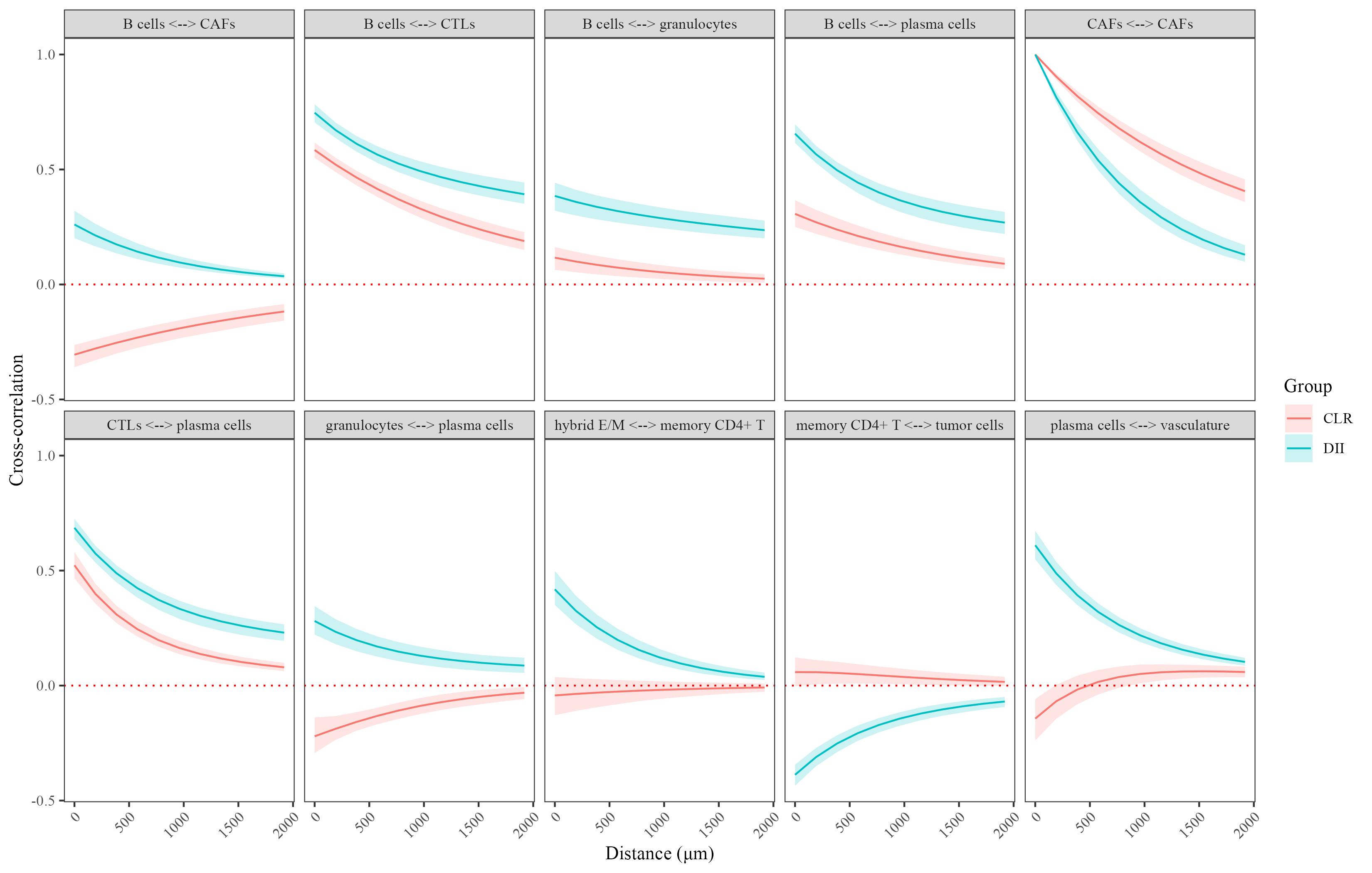}
\caption{The estimated spatial cross-correlations between select pairs of cell-type combinations from patients in the CLR group and the DII group. The colored line in each plot indicates the posterior expectation, while the shaded region represents the pointwise 95\% credible interval.}
\label{fig:xcor_CRC}
\vspace{1mm}
\end{sidewaysfigure}

The spatial cross-correlation analysis reveals distinct organizational patterns in the tumor microenvironment (TME) between CLR and DII patients, reflecting their respective immune-hot and immune-cold phenotypes (Figure~\ref{fig:xcor_CRC}). These differences highlight how immune and stromal cells are distributed differently across patient groups, with CLR tumors exhibiting a more spatially intermixed immune landscape and DII tumors displaying stronger compartmentalization, consistent with an immune-excluded state~\citep{fu_spatial_2021}.

\paragraph{Immune Exclusion and Stromal Structuring in DII Tumors}  

DII tumors exhibit significantly stronger spatial correlations between plasma cells and vasculature, suggesting that humoral immune responses in these tumors are spatially restricted to perivascular niches. This pattern indicates that plasma cells in DII tumors preferentially accumulate near blood vessels rather than distributing more broadly throughout the tumor microenvironment. Such confinement could limit their interactions with effector immune cells and tumor cells, reinforcing a segregated immune landscape \citep{nowosad_perivascular_2023}. Similarly, B cells show a much stronger association with cancer-associated fibroblasts (CAFs) in DII tumors, whereas in CLR tumors, this association is exclusionary. This suggests that CAFs in DII patients may be spatially structuring B cell responses, potentially forming barriers that restrict immune infiltration into tumor-rich regions \citep{mhaidly_role_2021}.

Granulocytes in DII tumors also exhibit a stronger correlation with plasma cells and B cells, which may indicate spatial compartmentalization of humoral immunity and myeloid-driven immune suppression. Given the known tumor-supportive roles of tumor-associated neutrophils (TANs), this spatial pattern could reflect an immune landscape where granulocytes are suppressing adaptive immune responses rather than promoting direct anti-tumor activity \citep{wu_tumor-associated_2020}. Hybrid epithelial-mesenchymal (E/M) cells in DII tumors also show significantly stronger correlations with memory CD4+ T cells, a pattern that may indicate localized immune suppression where hybrid E/M cells contribute to immune evasion by modifying their surrounding microenvironment \citep{fedele_permissive_2020}.

Notably, cytotoxic T lymphocytes (CTLs) exhibit stronger spatial correlations with plasma cells and B cells in DII tumors compared to CLR tumors. This suggests that in the immune-excluded environment of DII tumors, CTLs are more confined to lymphoid aggregates or peritumoral regions rather than being evenly distributed across the tumor microenvironment. This restricted pattern may reflect an immune response that is less capable of directly engaging tumor cells, reinforcing the immune-excluded phenotype of DII tumors~\citep{kinker_b_2021,wortman_spatial_2021}.

\paragraph{More Spatial Mixing and Less Exclusion in CLR Tumors}  

In contrast to the spatial segregation observed in DII tumors, CLR tumors exhibit a more intermixed immune and stromal landscape, where spatial correlations between cell types are generally closer to zero. This suggests that CLR tumors lack the pronounced immune compartmentalization seen in DII tumors, enabling more uniform mixing of immune and stromal populations. For example, memory CD4+ T cells in CLR tumors exhibit weaker spatial correlation with vasculature and CAFs compared to DII tumors, suggesting that helper T cells are more evenly distributed throughout the tumor rather than being concentrated in specific stromal niches. This well-mixed distribution could enhance immune surveillance and support cytotoxic responses in CLR patients \citep{baker_cancer-associated_2021,chen_cafs_2024}. 

In contrast to most other stromal-immune associations, CAFs in CLR tumors exhibit a notably stronger self-association than in DII tumors, suggesting that fibroblasts in CLR patients are more spatially structured. This pattern may indicate that certain CAF populations in CLR tumors play an active role in reinforcing immune-permissive microenvironments, potentially supporting tertiary lymphoid structure (TLS) formation or immune cell recruitment \citep{chen_cafs_2024, gunaydin_cafs_2021,guo_cancer-associated_2023}. Rather than forming exclusively physical barriers, as they appear to in DII tumors, CAFs in CLR tumors may contribute to organized immune-stromal interactions that facilitate adaptive immune responses.

Interestingly, while CLR tumors generally do not exhibit significantly stronger immune cell interactions in the way that DII tumors do, they show less exclusionary spatial patterns between immune and tumor cells. For instance, memory CD4+ T cells exhibit a much less negative spatial correlation with both tumor cells and hybrid E/M cells in CLR tumors, suggesting that helper T cells are not being actively excluded from tumor-rich regions to the same extent as in DII tumors. Rather than forming segregated zones, the immune and tumor compartments in CLR tumors appear to be more diffusely distributed, supporting a microenvironment where immune cells may have better access to engage tumor cells directly.

\paragraph{Conclusion}  

The spatial correlation patterns between immune and stromal components highlight the fundamental differences between CLR and DII tumors. While DII tumors exhibit strong compartmentalization, particularly through structured immune-stromal associations, CLR tumors show a more intermixed spatial organization where immune and tumor cells are less strictly segregated. This suggests that CLR tumors may support greater immune cell mobility and accessibility to tumor regions, potentially facilitating more effective anti-tumor immune responses.

\section{Discussion}\label{sc:discuss}

In this work, we developed a novel Bayesian hierarchical framework for jointly modeling spatial dependencies in multiplexed imaging data across multiple subjects. 
Our methods provide a principled statistical approach to pool information across patients and infer cell-type-specific spatial co-location patterns. The interpretable output from our model is a set of disease-specific spatial correlation functions that quantify how strongly different cell types associate spatially over varying distances. We applied our methods to UM-PCC and CRC data, revealing novel insights into immune infiltration patterns, immune suppression, TME organization, and the role of spatial organization in cancer progression and immune evasion. 

Specifically, on pancreatic cancer data, our methods revealed a stark contrast in immune cell spatial patterns between IPMN and PDAC patients. In IPMN, a more well-mixed immune microenvironment suggests a relatively intact immune surveillance system. Conversely, in PDAC, immune exclusion was evident through the spatial segregation of T cells together away from tumor cells. This supports the hypothesis that tumor progression is associated with restructuring of the immune microenvironment to evade immune detection~\citep{jamouss_tumor_2025}. Such findings align with prior work suggesting that PDAC fosters an immunosuppressive niche that inhibits effective anti-tumor responses~\citep{hernandez_diminished_2022}. Similarly, our analysis of colorectal cancer (CRC) data highlighted significant differences between the Crohn’s-like reaction (CLR) and diffuse inflammatory infiltration (DII) patient groups. CLR patients exhibited a more integrated immune microenvironment, consistent with the presence of tertiary lymphoid structures and enhanced immune surveillance~\citep{maoz_crohns-like_2019}. In contrast, the observed interactions between cancer-associated fibroblasts (CAFs) and immune cells in DII patients further underscored the complex crosstalk shaping tumor immune landscapes, particularly in promoting immune suppression and tumor progression~\citep{dang_cancer-associated_2023}. 

Future research will address several open challenges. First, our approach assumes a common spatial covariance structure across all subjects within a given disease group. While this assumption facilitates information sharing, it may not fully capture inter-patient heterogeneity. Future work could explore hierarchical extensions that allow partial pooling of spatial parameters, better accommodating variations in spatial dependencies across individuals. Methodological advancements in this direction might build on recent developments in Bayesian factor modeling with multi-study data \citep{de_vito_multi-study_2019, mauri_spectral_2025}.
Second, our model assumes that the spatial patterns of all cell types can be represented via the same set of spatial factors, leading to inflexibility in modeling cell-type-specific spatial characteristics. Recent developments in more flexible cross-covariance function models \citep{peruzzi_inside-out_2024} may help resolve these limitations.
Third, our model can be extended via integration of covariates, such as patient-specific information, directly into the covariance model. Covariate-dependent spatial correlation curves have potential implications for therapeutic targeting. 

Finally, our approach is generally applicable to other spatially resolved omics technologies, such as spatial transcriptomics or mass cytometry imaging. These modalities provide complementary information about gene expression and protein abundance, enabling a multi-modal characterization of the tumor microenvironment.

\newpage
\begin{center}
  {\large\bf SUPPLEMENTARY MATERIAL}
\end{center}
\appendix

\section{Simulation Details}
\label{apdx:sims}
Both sets of simulations used the following sets of equations to simulate from this model: 
\begin{align*}
    \varphi_j &\sim \text{Uniform}(\phi_{\min}, \phi_{\max}), \quad j = 1, \dots, k \\
    L_j &= \text{Cholesky} \left( \exp(-\varphi_j D) \right), \quad j = 1, \dots, k \\
    \quad \varepsilon_{i,j} &\sim \mathcal{N}(0, I_{n}), \quad i = 1, \dots, N , \quad j = 1, \dots, k\\
    V_i &= \begin{bmatrix} L_1 \varepsilon_{i,1} & L_2 \varepsilon_{i,2} & \dots & L_k \varepsilon_{i,k} \end{bmatrix} \\
    \beta_0 &= \balpha\mathbf{1}_{1 \times q} \\
    X_i &= \mathbf{1}_{n \times 1}, \quad i = 1, \dots, N \\
    W_{i} &= V_i \bA^T + X_i \beta_0, \quad i = 1, \dots, N \\
    \by_i &\sim \text{Poisson} \left( \exp(W_i) \right), \quad i = 1, \dots, N
\end{align*}
\begin{align*}
\bA &= \begin{bmatrix} 
\lambda_{11} & 0 & 0 & \dots & 0 \\ 
\lambda_{21} & \lambda_{22} & 0 & \dots & 0 \\ 
\vdots & \vdots & \ddots & \dots & 0 \\ 
\lambda_{k1} & \lambda_{k2} & \dots & \dots & \lambda_{kk} \\
\vdots & \vdots & \ddots & \dots & \vdots \\ 
\lambda_{q1} & \lambda_{q2} & \dots & \dots & \lambda_{qk} \\
\end{bmatrix} 
\end{align*}

where:
\begin{itemize}
    \item $q$ is the number of simulated cell types, $k$ is the number of latent factors, $N$ is the number of images to be simulated and $n$ is the total number of grid pixels
    \item \( \lambda_{ii} \sim \text{Uniform}(0.5,1) \) for diagonal elements and \( \lambda_{ij} \sim \text{Uniform}(-0.7, 0.7) \) for below-diagonal elements.
    \item  $D\in \mathbb{R}^{n\times n}$ is the pairwise distance matrix.
    \item $L_j\in \mathbb{R}^{n\times n}$ are the Cholesky factors of the covariance matrices.
    \item $V_i\in \mathbb{R}^{n\times k}$ are the latent factor matrices.
    \item $\bA\in \mathbb{R}^{q\times k}$ is the factor loading matrix.
    \item \( X_i \) is the design matrix.
    \item \( \beta_0 \) is the intercept term representing the baseline intensity offset.
    \item $W_i\in \mathbb{R}^{n\times q}$ is the transformed latent space.
    \item $\by_i\in \mathbb{R}^{n\times q}$ is the observed count data modeled via a Poisson likelihood.
\end{itemize}

\section{Supplemental Results}
\subsection{Recovering Cross-Correlations When the Number of Latent Factors is Misspecified}

In this section, we examine the impact of misspecifying the number of latent factors, $k$, on the estimation of cross-correlation functions. The true number of latent factors is denoted as $k^* = 3$, and we compare cases where $k$ is under-specified ($k=2$), correctly specified ($k=3$), and over-specified ($k=6$). We also analyze the stability and reliability of these estimates using metrics such as effective sample size (ESS), Gelman-Rubin diagnostic ($\hat{R}$), and Widely Applicable Information Criterion (WAIC).

\begin{figure}[h]
\centering
\includegraphics[width=\textwidth]{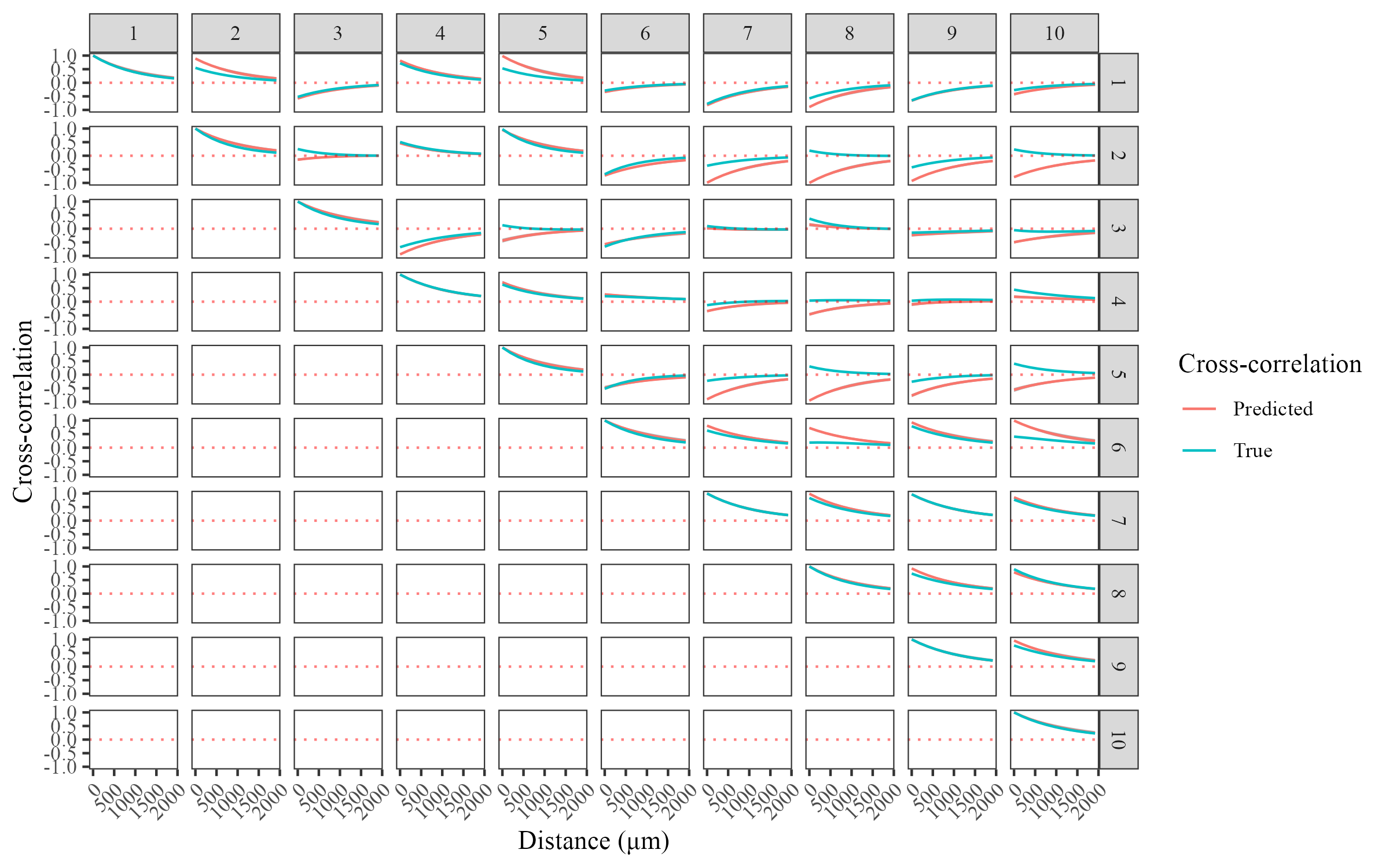}
\caption{An example of cross-correlation curves estimated using $k=2$, when $k^*=3$.}
\label{fig:cross_cor_sim1_k2}
\vspace{1mm}
\end{figure}

\begin{figure}[h]
\centering
\includegraphics[width=\textwidth]{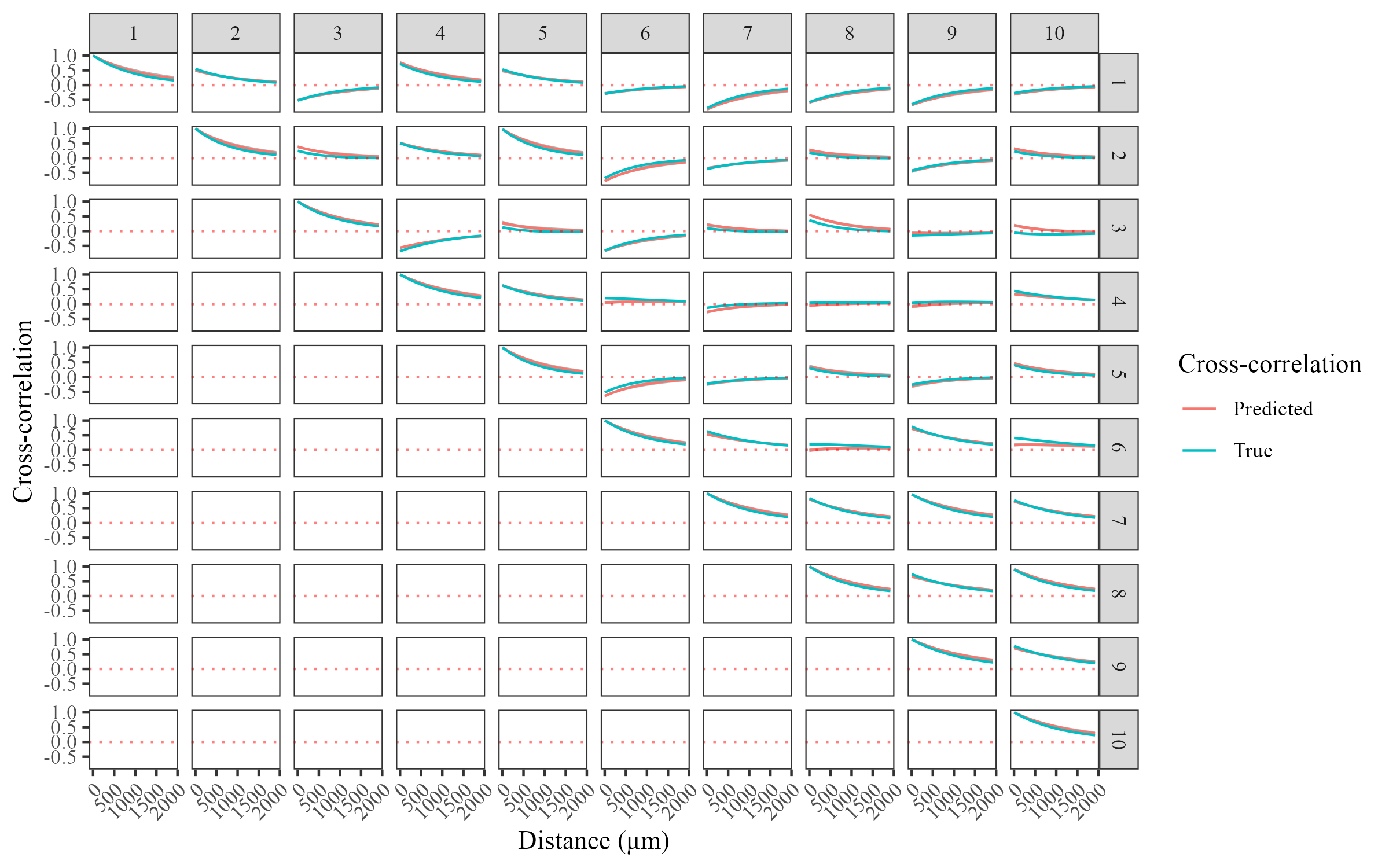}
\caption{An example of cross-correlation curves estimated using $k=3$, when $k^*=3$.}
\label{fig:cross_cor_sim2_k3}
\vspace{1mm}
\end{figure}

\begin{figure}[h]
\centering
\includegraphics[width=\textwidth]{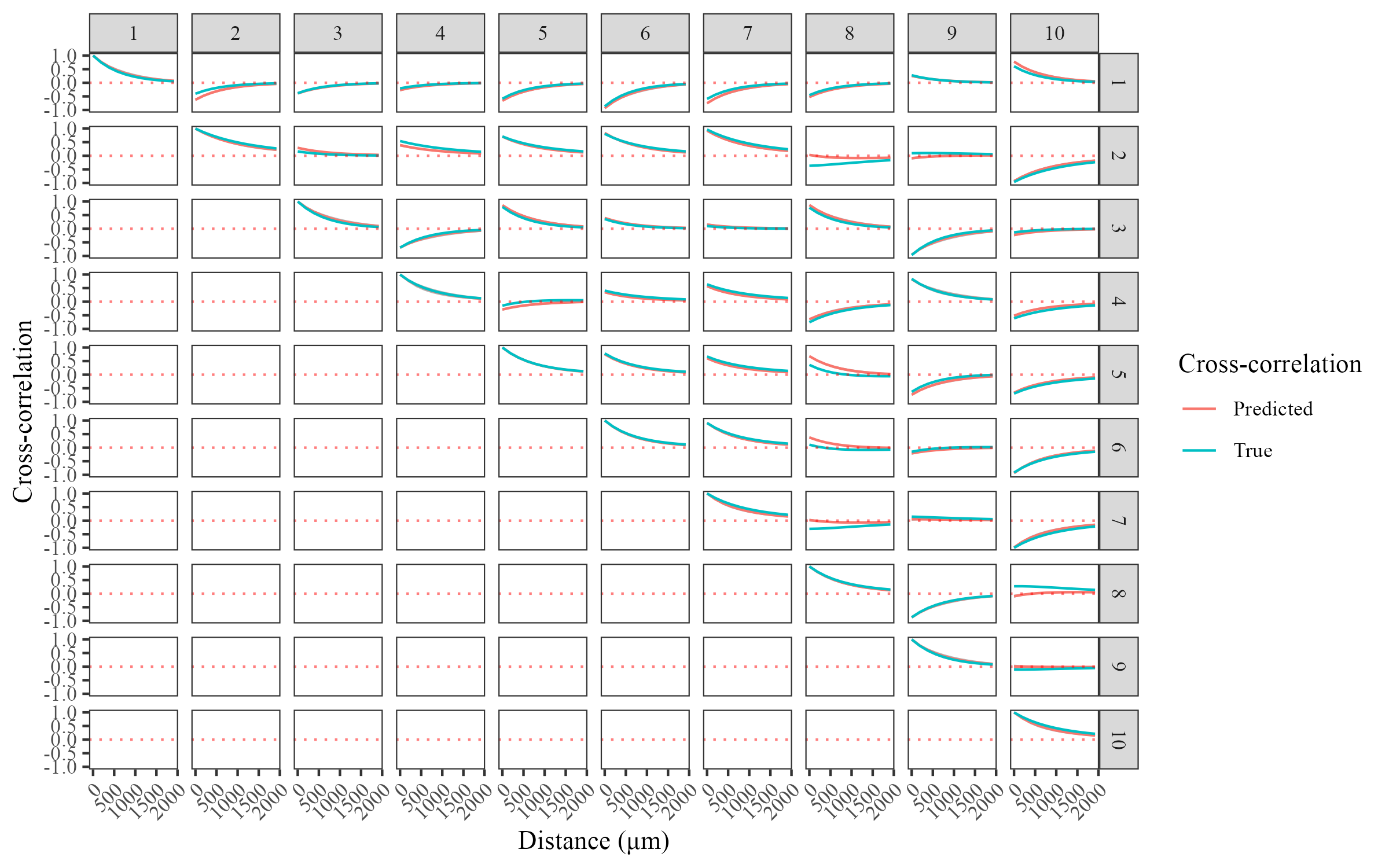}
\caption{An example of cross-correlation curves estimated using $k=6$, when $k^*=3$.}
\label{fig:cross_cor_sim9_k6}
\vspace{1mm}
\end{figure}

Figure \ref{fig:cross_cor_sim1_k2} shows an example of estimated cross-correlation curves when the number of latent factors is under-specified ($k=2$). Compared to the true model ($k^*=3$), the estimates exhibit large deviations for many pairs, indicating underfitting. In contrast, Figure \ref{fig:cross_cor_sim2_k3} presents the cross-correlation estimates for the correctly specified model ($k=3$), demonstrating improved recovery of the underlying spatial dependence structure. When the model is over-specified ($k=6$), as shown in Figure \ref{fig:cross_cor_sim9_k6}, the cross-correlation estimates remain largely stable but exhibit slight increases in deviations from the true curves.

\begin{figure}[h]
\centering
\includegraphics[width=\textwidth]{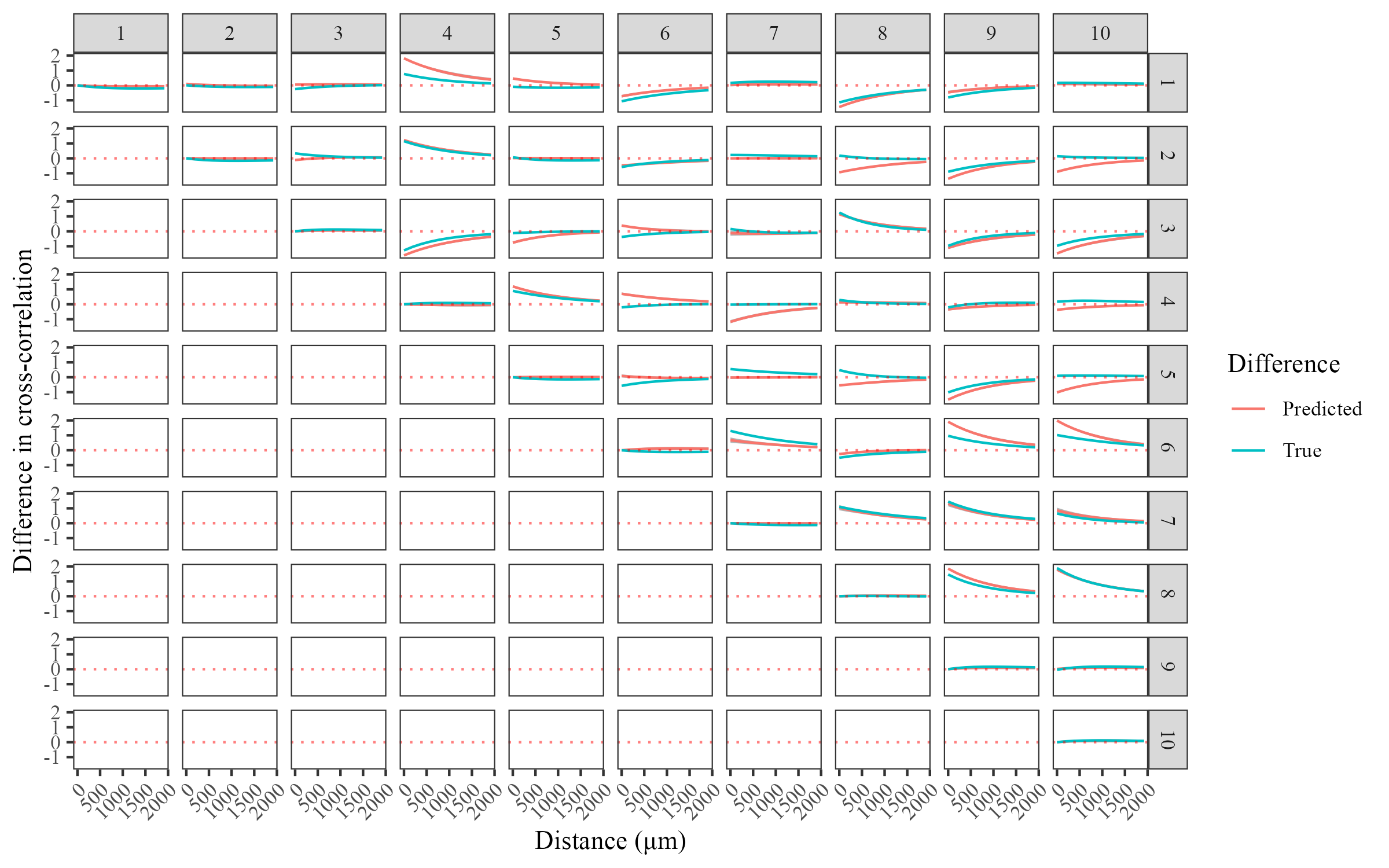}
\caption{An example of the difference in two sets of cross-correlation curves that were both estimated using $k=2$, when $k^*=3$.}
\label{fig:group_diff_k}
\vspace{1mm}
\end{figure}

To further assess estimation differences, Figure \ref{fig:group_diff_k} visualizes the difference in two sets of cross-correlation curves estimated using $k=2$. This comparison highlights the degree of variability introduced when the number of latent factors is underspecified.

\begin{figure}[h]
\centering
\includegraphics[width=\textwidth]{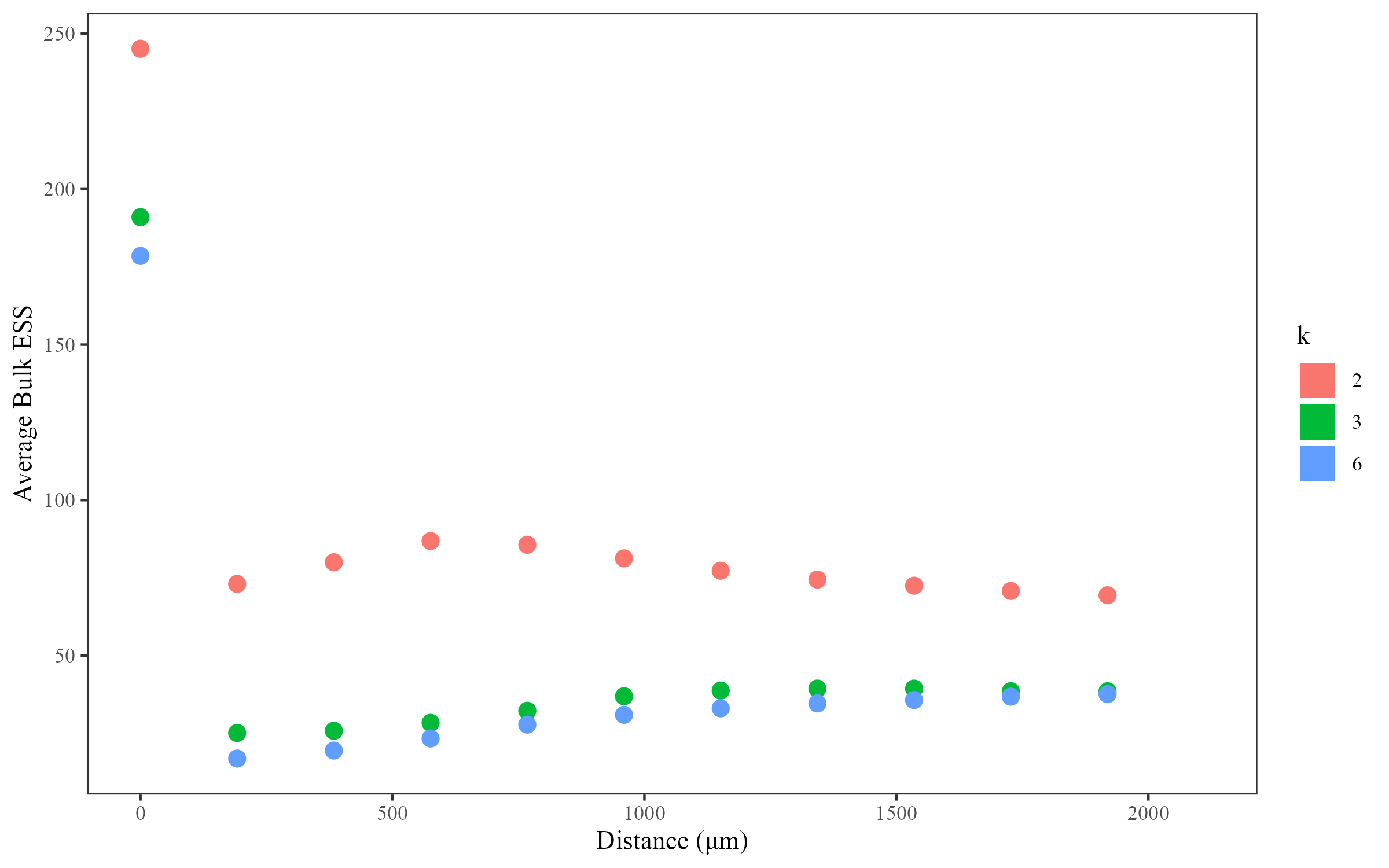}
\caption{Effective sample sizes (ESS) of cross-correlation estimates at different spatial distances, over varying $k$.}
\label{fig:bulk_ess_k}
\vspace{1mm}
\end{figure}

\begin{figure}[h]
\centering
\includegraphics[width=\textwidth]{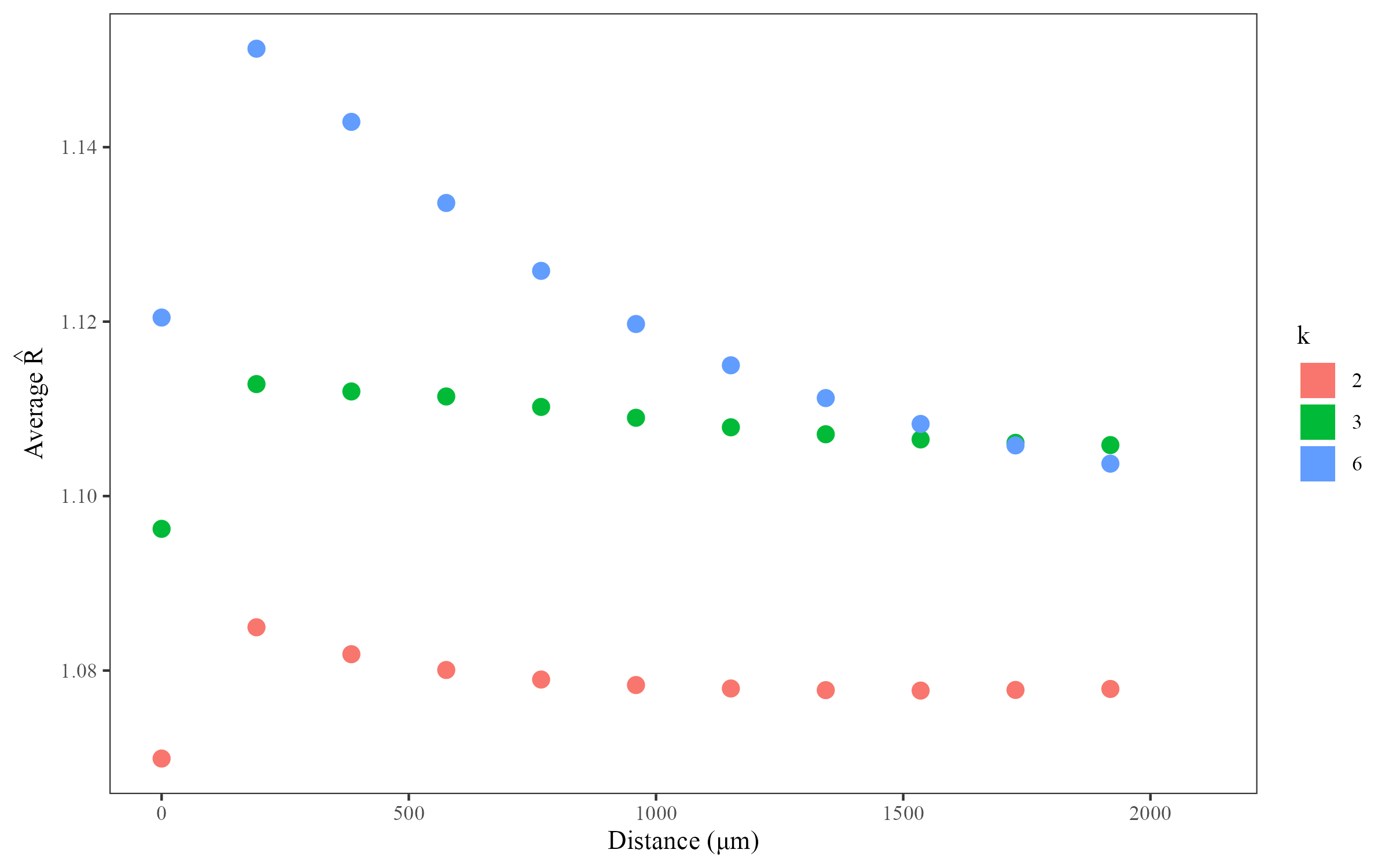}
\caption{$\hat{R}$ of cross-correlation estimates at different spatial distances, over varying $k$.}
\label{fig:rhat_k}
\vspace{1mm}
\end{figure}

We evaluate the efficiency of cross-correlation estimates by examining their effective sample sizes (ESS) at different spatial distances (Figure \ref{fig:bulk_ess_k}). The ESS provides insight into the quality of posterior samples, with lower values indicating higher autocorrelation and potential inefficiencies in MCMC sampling. Additionally, Figure \ref{fig:rhat_k} presents the Gelman-Rubin diagnostic ($\hat{R}$) across varying $k$, where values close to 1 indicate convergence.

\begin{figure}[h]
\centering
\includegraphics[width=\textwidth]{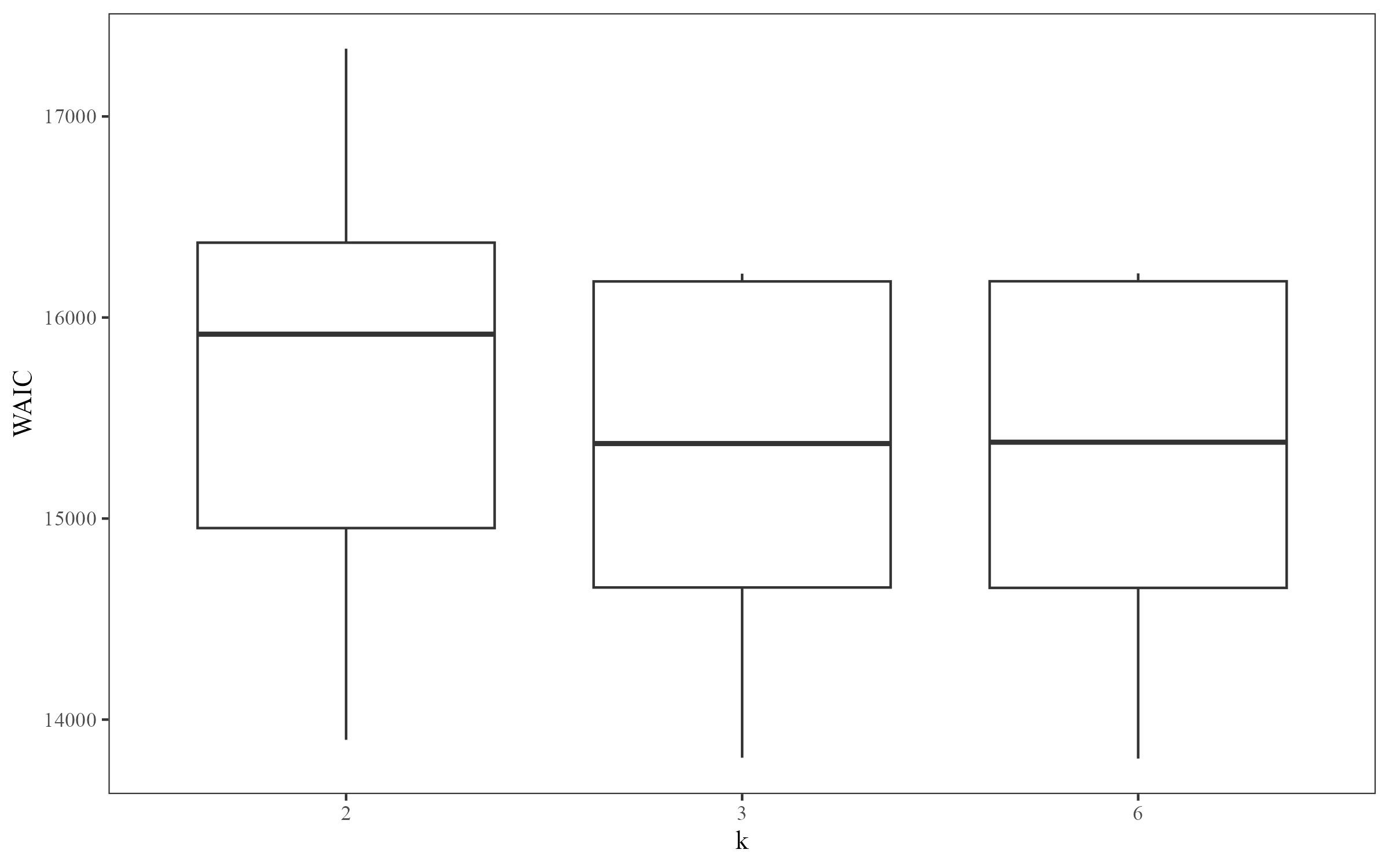}
\caption{The distribution of WAIC for different $k$, when $k^*$=3.}
\label{fig:waic_k}
\vspace{1mm}
\end{figure}

To assess model fit, Figure \ref{fig:waic_k} shows the distribution of WAIC values across different choices of $k$. The lowest WAIC values correspond to the correctly specified model ($k=3$), supporting its selection as the optimal choice. However, models with $k=6$ have WAIC values very close to those from models with $k=3$.

\begin{figure}
\centering
\includegraphics[width=\textwidth]{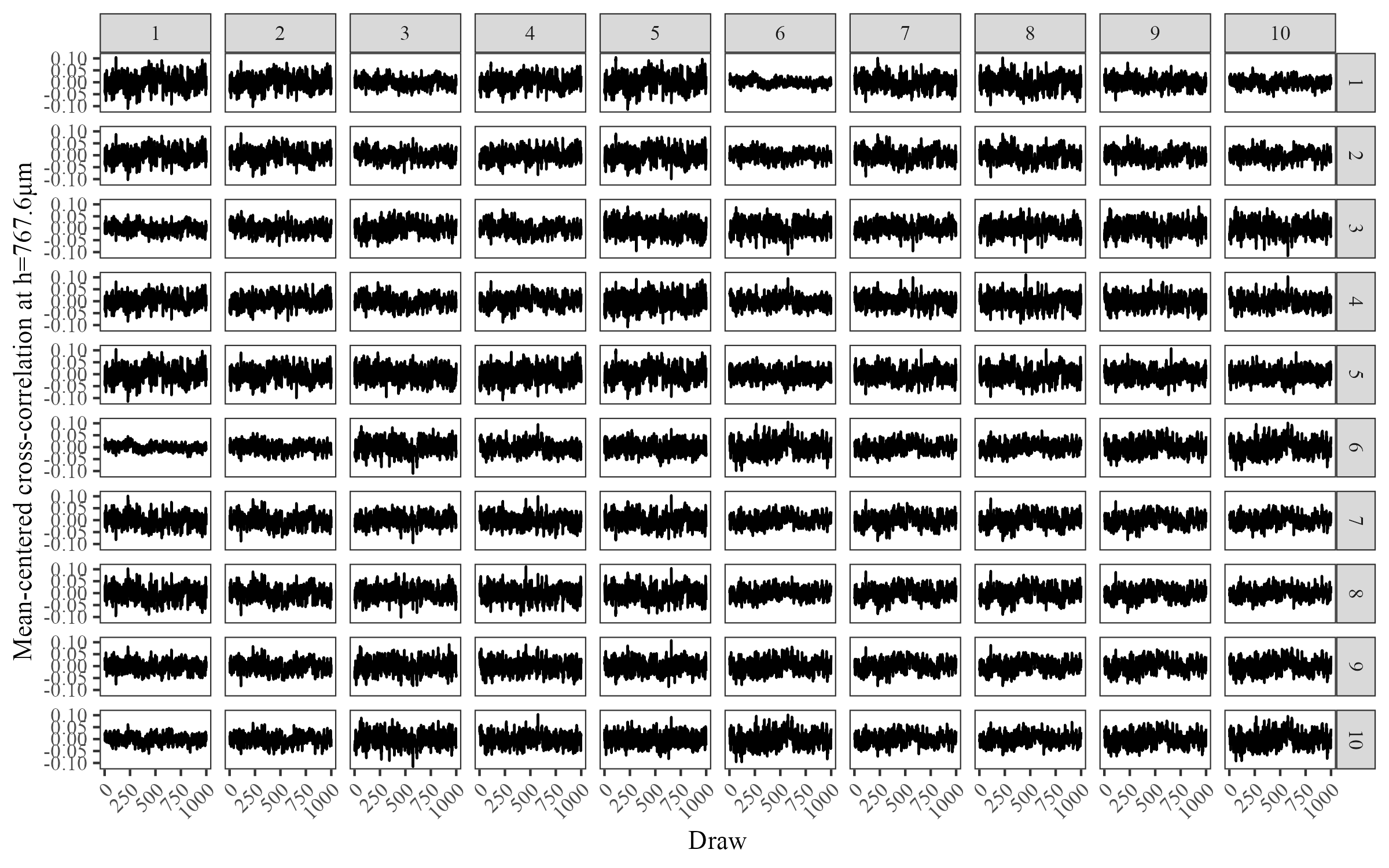}
\caption{Example of MCMC trace plot for estimated mean-centered cross-correlations at $h=767.6\mu m$, using $k=2$.}
\label{fig:trace_df_k2}
\vspace{1mm}
\end{figure}

Next, Figure \ref{fig:trace_df_k2} provides an example of an MCMC trace plot for estimated mean-centered cross-correlations at a spatial distance of $h=767.6 \mu m$, when $k=2$. The trace plot reveals the mixing behavior of the MCMC chains, which is crucial for ensuring robust inference.

\begin{figure}[h]
\centering
\includegraphics[width=\textwidth]{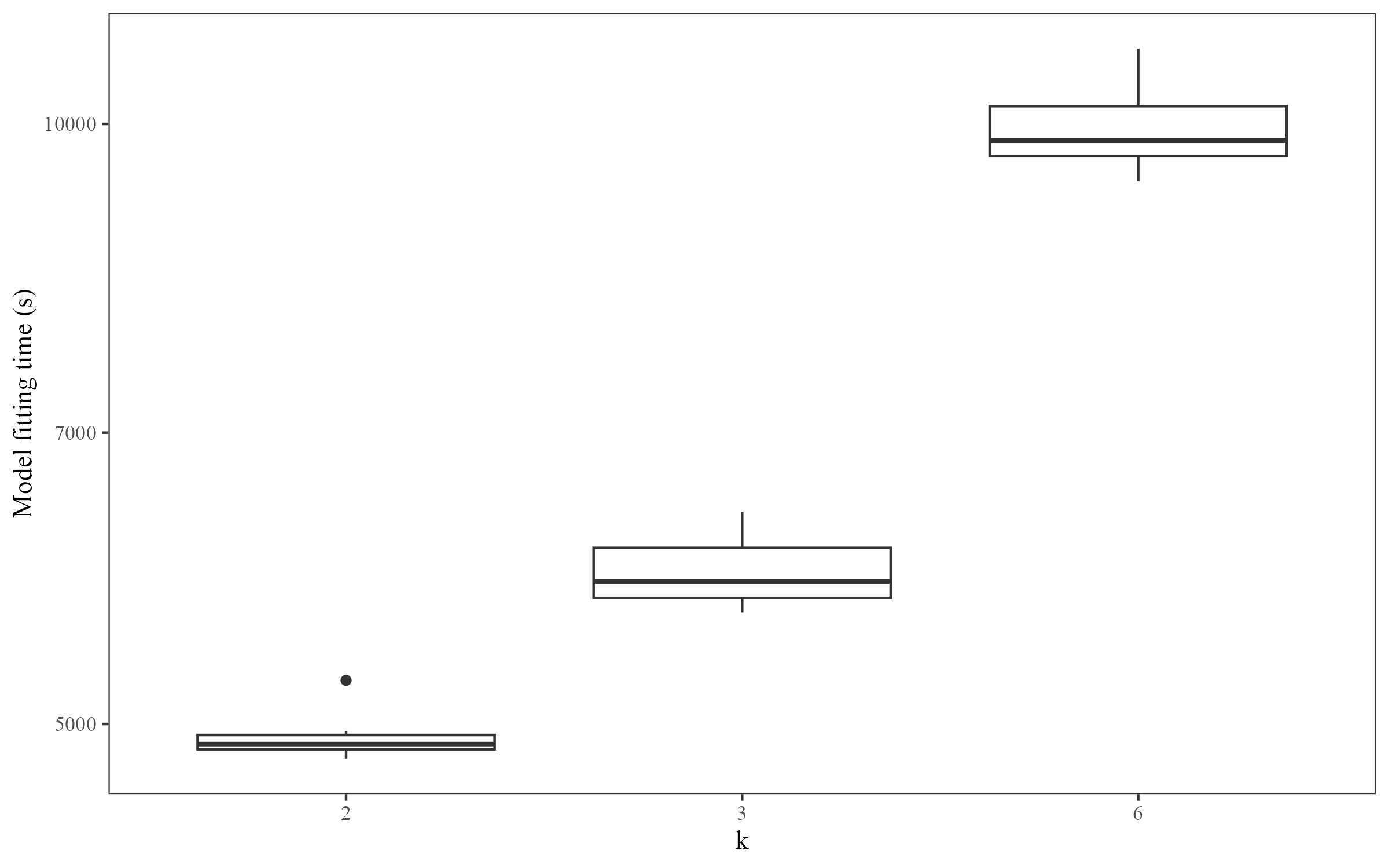}
\caption{Model fitting times for simulations at various $k$.}
\label{fig:timings_k}
\vspace{1mm}
\end{figure}

Finally, we show the model fitting times for various $k$ in Figure \ref{fig:timings_k}, demonstrating the linear increase in fitting time as $k$ increases.

These results collectively illustrate how misspecifying $k$ influences cross-correlation estimation and highlight the importance of model selection criteria such as WAIC in guiding the choice of an appropriate number of latent factors.

\clearpage
\subsection{The Effect of Grid Size on Cross-Correlation Recovery}

In this section, we assess how the choice of grid size, $n_x$, affects the recovery of cross-correlation functions. The grid size determines the resolution of spatial intensity estimates, impacting both estimation accuracy and computational efficiency. We consider three different grid sizes: $n_x \in {12, 24, 48}$.

\begin{figure}[h]
\centering
\includegraphics[width=\textwidth]{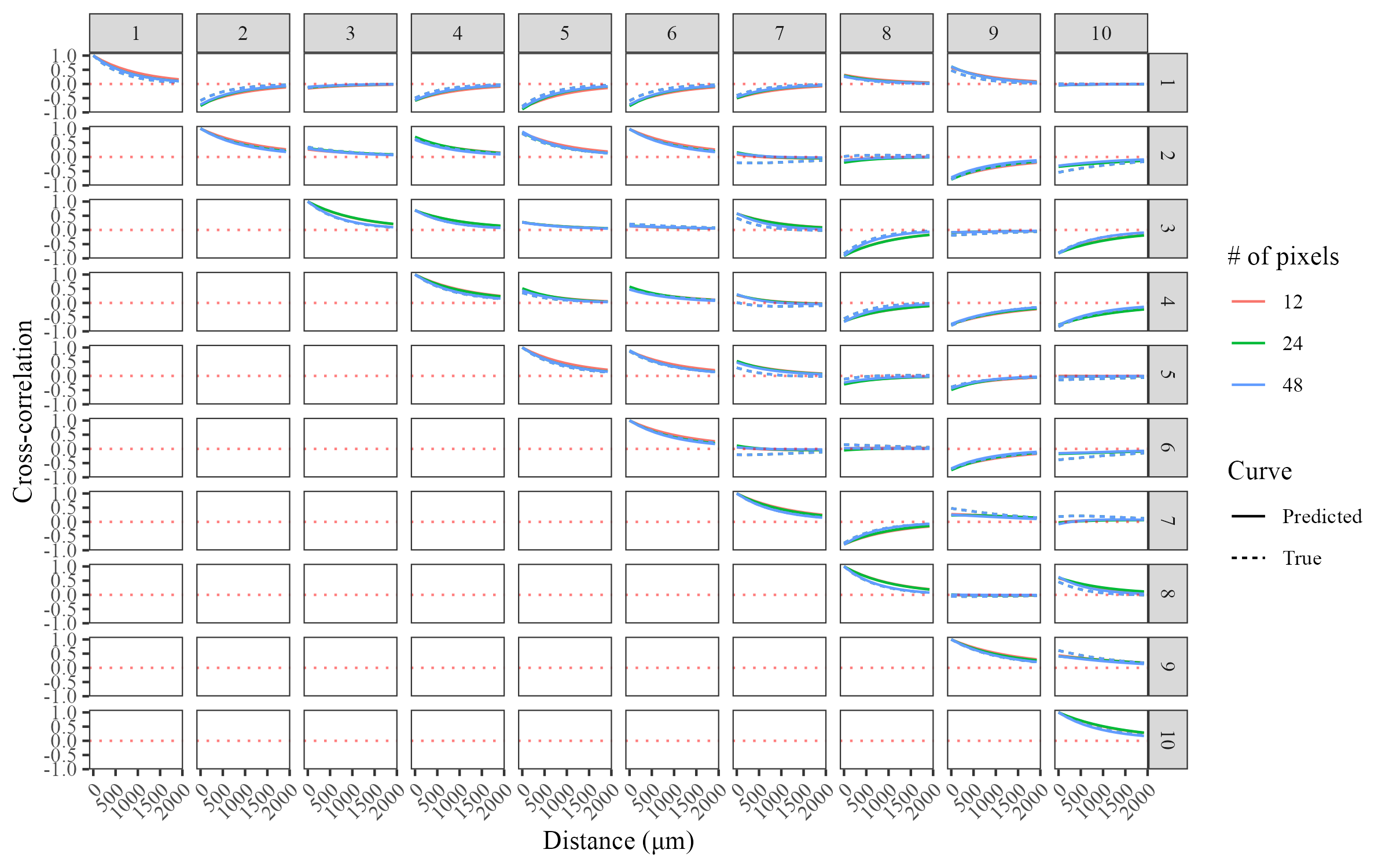}
\caption{A comparison of cross-correlation curves estimated using $n_{x}\in\{12,24,48\}$}
\label{fig:cross_cor_compare_sz}
\vspace{1mm}
\end{figure}

Figure \ref{fig:cross_cor_compare_sz} compares the estimated cross-correlation curves for these different grid sizes. The results demonstrate that there are not large differences in estimated cross-correlation curves across various values of $n_x$.

\begin{figure}[h]
\centering
\includegraphics[width=\textwidth]{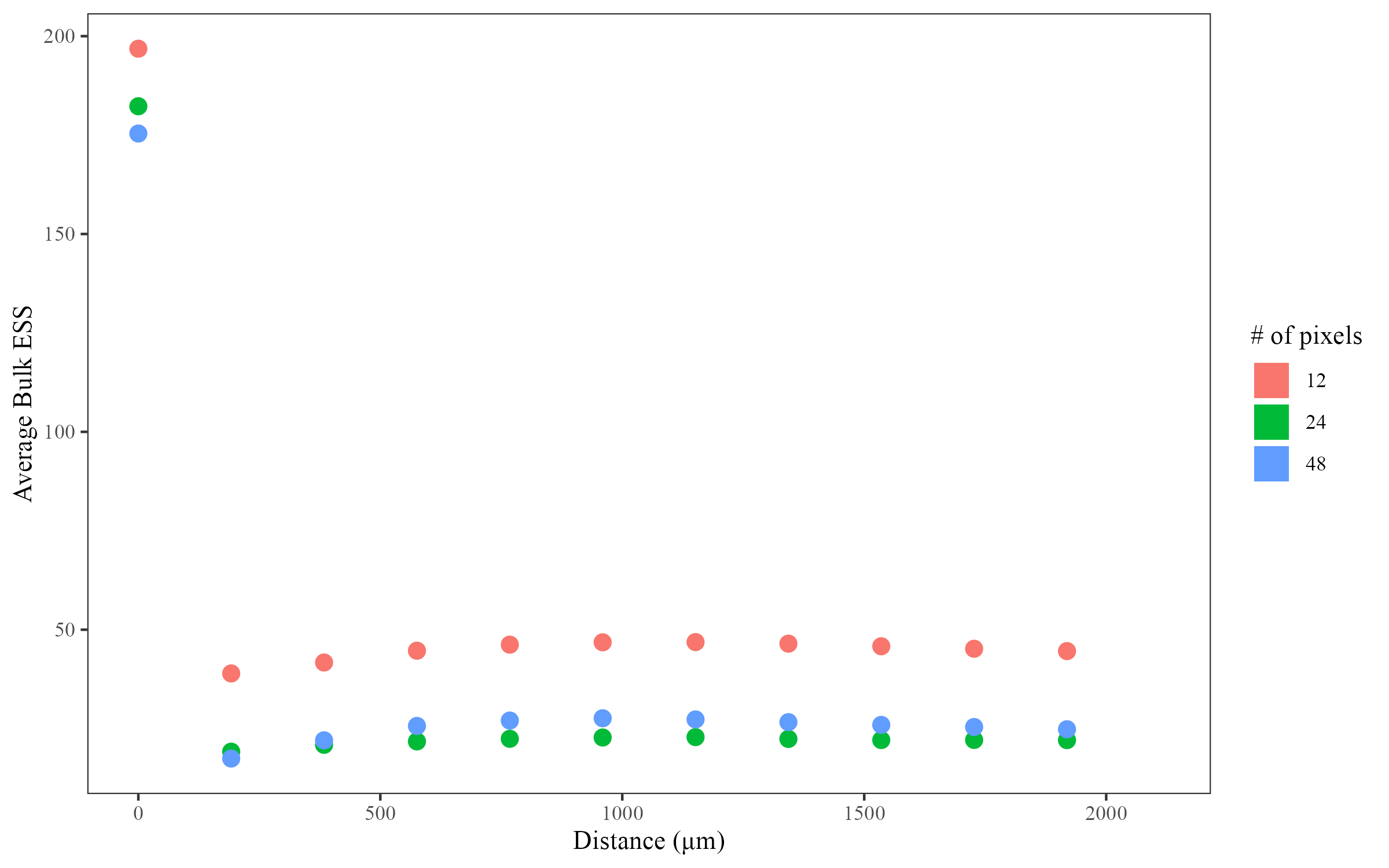}
\caption{Effective sample sizes (ESS) of cross-correlation estimates at different spatial distances, over varying $n_{x}$.}
\label{fig:bulk_ess_sz}
\vspace{1mm}
\end{figure}

To evaluate sampling efficiency, Figure \ref{fig:bulk_ess_sz} presents the effective sample sizes (ESS) for different spatial distances across varying $n_x$. Lower grid resolutions tend to yield larger ESS values, suggesting more efficient MCMC sampling and reduced autocorrelation in posterior estimates.

\begin{figure}[h]
\centering
\includegraphics[width=\textwidth]{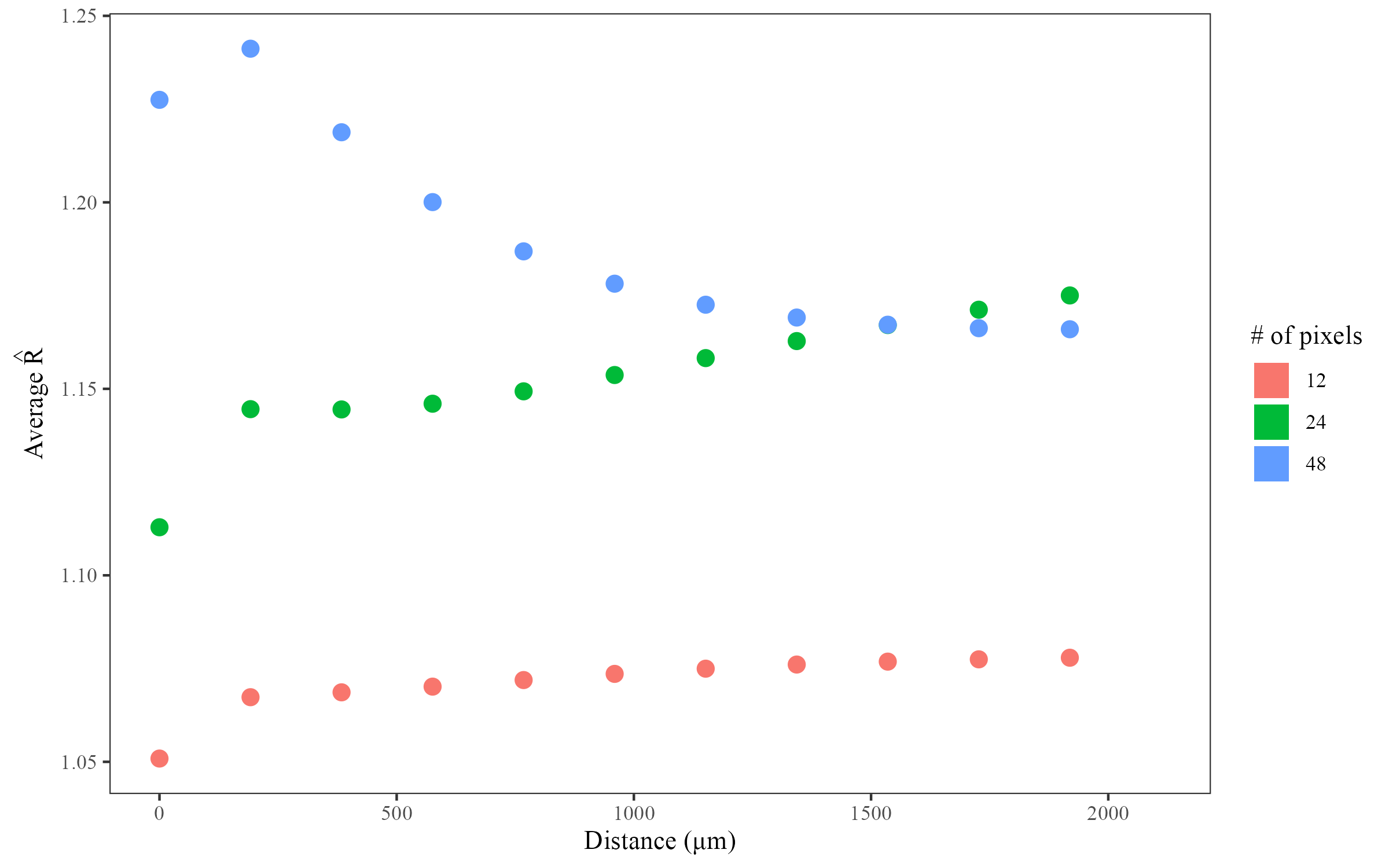}
\caption{$\hat{R}$ of cross-correlation estimates at different spatial distances, over varying $n_{x}$.}
\label{fig:rhat_sz}
\vspace{1mm}
\end{figure}

Convergence diagnostics are shown in Figure \ref{fig:rhat_sz}, which displays the Gelman-Rubin diagnostic ($\hat{R}$) across different grid sizes. Values of $\hat{R}$ close to 1 indicate satisfactory convergence, and we observe that coarser grids ($n_x=12$) maintain the best convergence properties.

\begin{figure}[h]
\centering
\includegraphics[width=\textwidth]{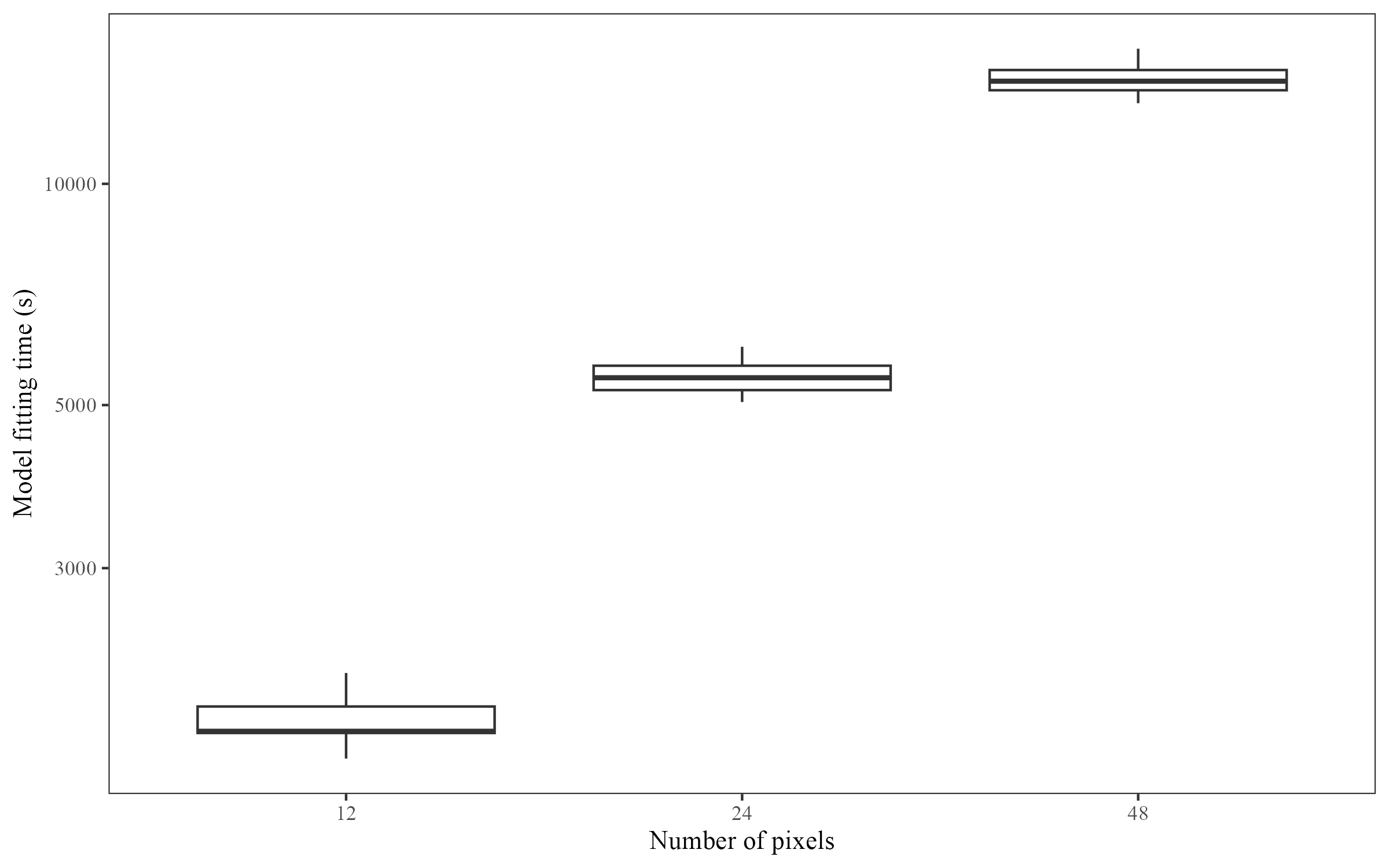}
\caption{Model fitting times for simulations at various $n_x$.}
\label{fig:timings_sz}
\vspace{1mm}
\end{figure}

Lastly, we show the model fitting times for various $n_x$ in Figure \ref{fig:timings_sz}, demonstrating the quadratic increase in fitting time as $n_x$ increases.

\clearpage
\subsection{Choosing $k$ for PANC Analysis}

In this section, we evaluate the selection of the number of latent factors, $k$, for pancreatic cancer (PANC) analysis by examining model convergence and the stability of cross-correlation estimates. We analyze two patient groups: PDAC (pancreatic ductal adenocarcinoma) and IPMN (intraductal papillary mucinous neoplasm).

Table \ref{tab:timings_panc} shows the model fitting times for various $k$ in both patient groups.

\begin{figure}[htb]
\centering
\includegraphics[width=\textwidth]{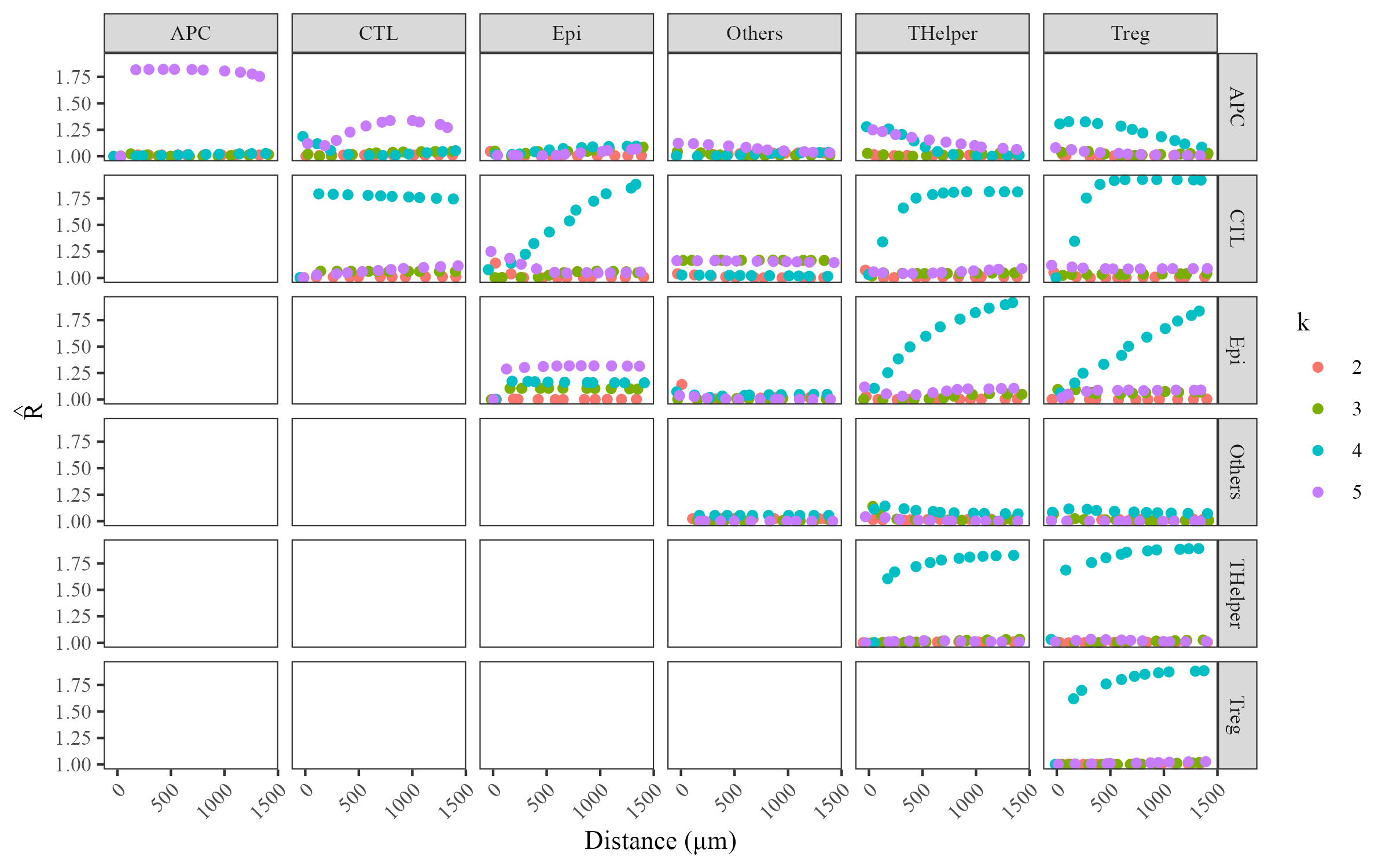}
\caption{$\hat{R}$ values to determine convergence of MCMC sampling of cross-correlations at various distances, under various $k$, in the PDAC patient group. Values closer to 1 indicate better convergence.}
\label{fig:panc_choosingk/rhat_1}
\vspace{1mm}
\end{figure}

\begin{figure}[htb]
\centering
\includegraphics[width=\textwidth]{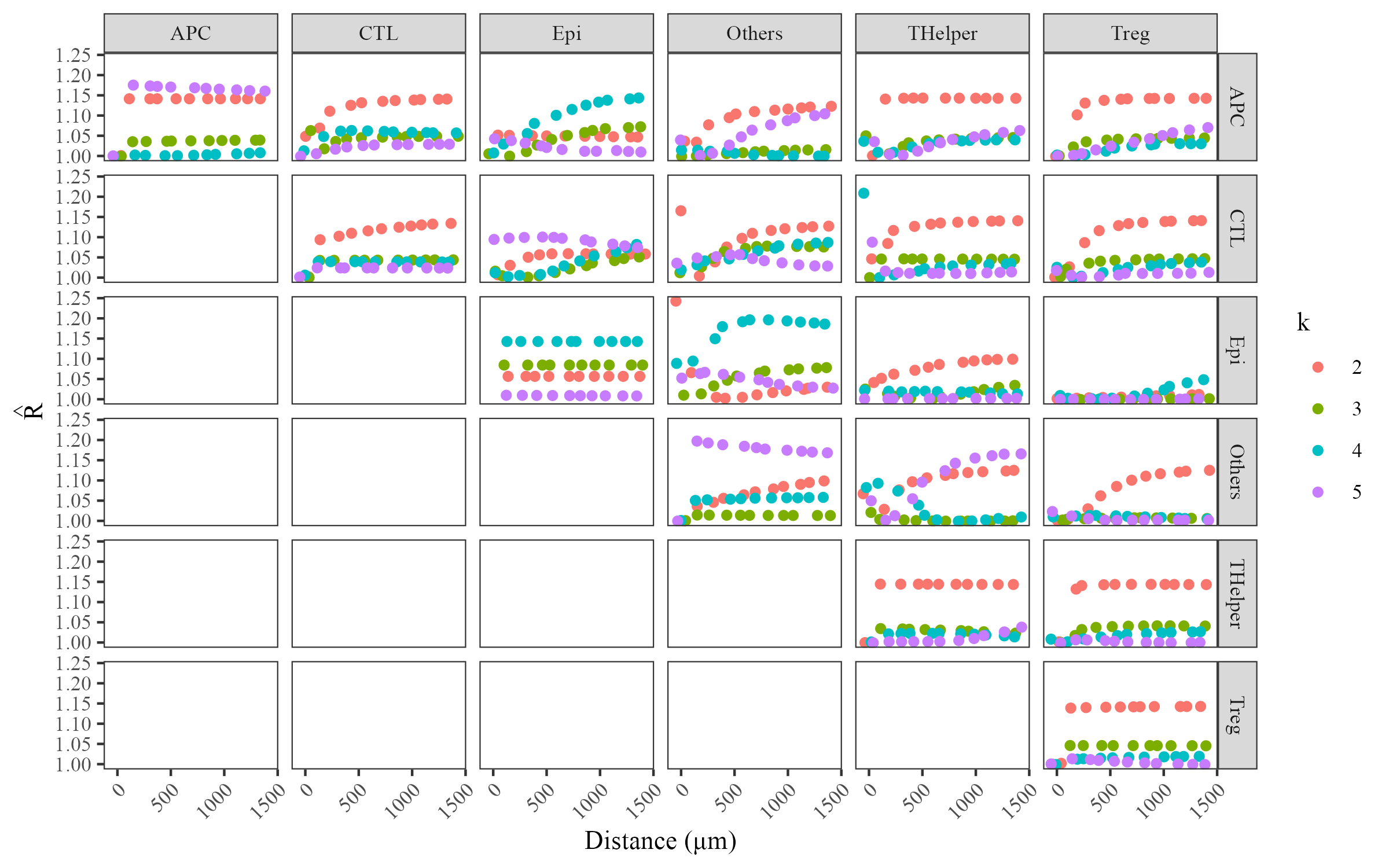}
\caption{$\hat{R}$ values to determine convergence of MCMC sampling of cross-correlations at various distances, under various $k$, in the IPMN patient group. Values closer to 1 indicate better convergence.}
\label{fig:panc_choosingk/rhat_2}
\vspace{1mm}
\end{figure}

Figures \ref{fig:panc_choosingk/rhat_1} and \ref{fig:panc_choosingk/rhat_2} display the Gelman-Rubin diagnostic ($\hat{R}$) values for cross-correlation estimates at different spatial distances across various $k$ in the PDAC and IPMN patient groups, respectively. Convergence is indicated by $\hat{R}$ values close to 1. Our results show that convergence results are mixed for various values of $k$.

\begin{figure}[htb]
\centering
\includegraphics[width=\textwidth]{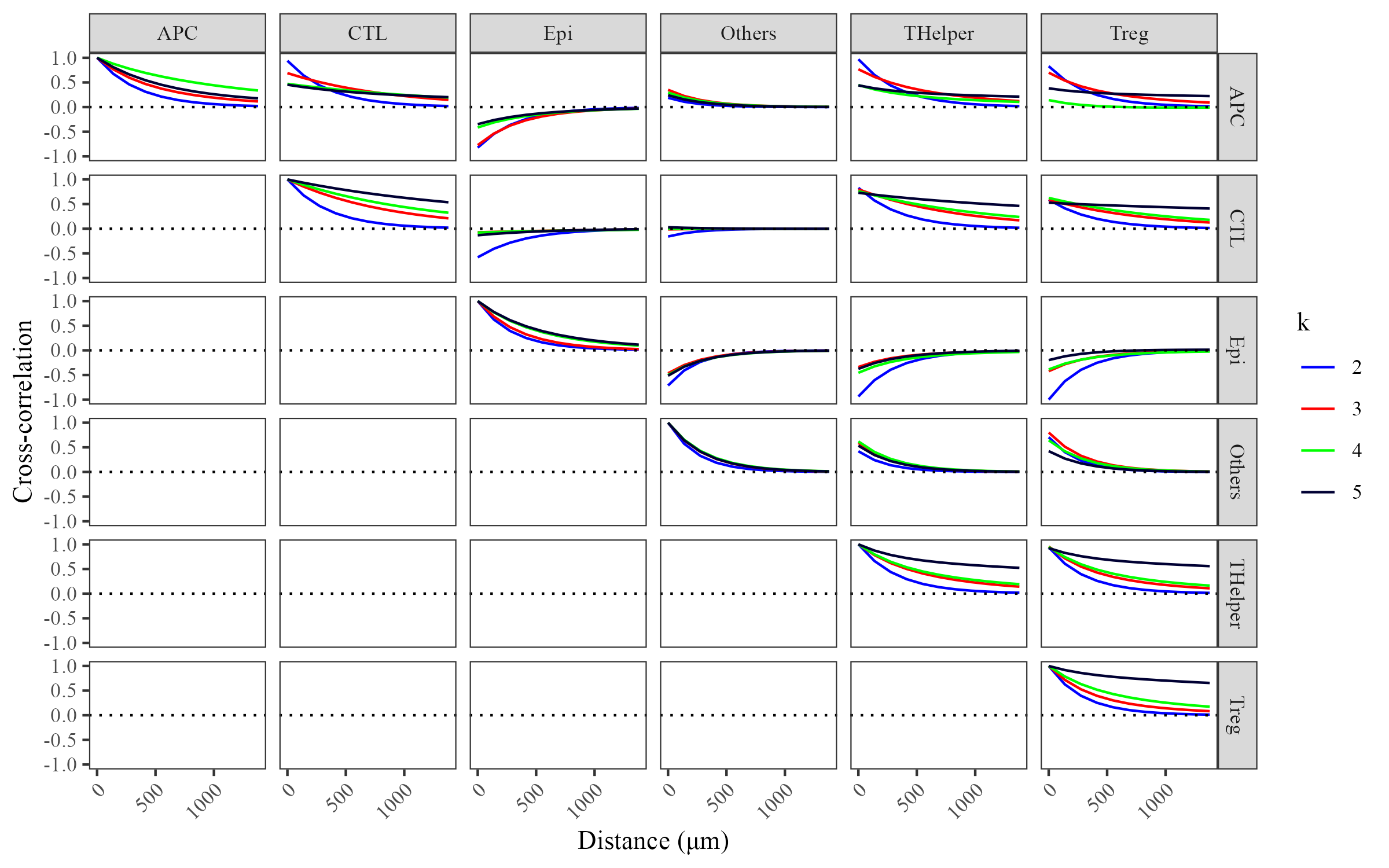}
\caption{Estimated cross-correlations at various distances, estimated at various $k$, in the PDAC patient group.}
\label{fig:panc_choosingk/xcor_varyingk_group1}
\vspace{1mm}
\end{figure}

\begin{figure}[htb]
\centering
\includegraphics[width=\textwidth]{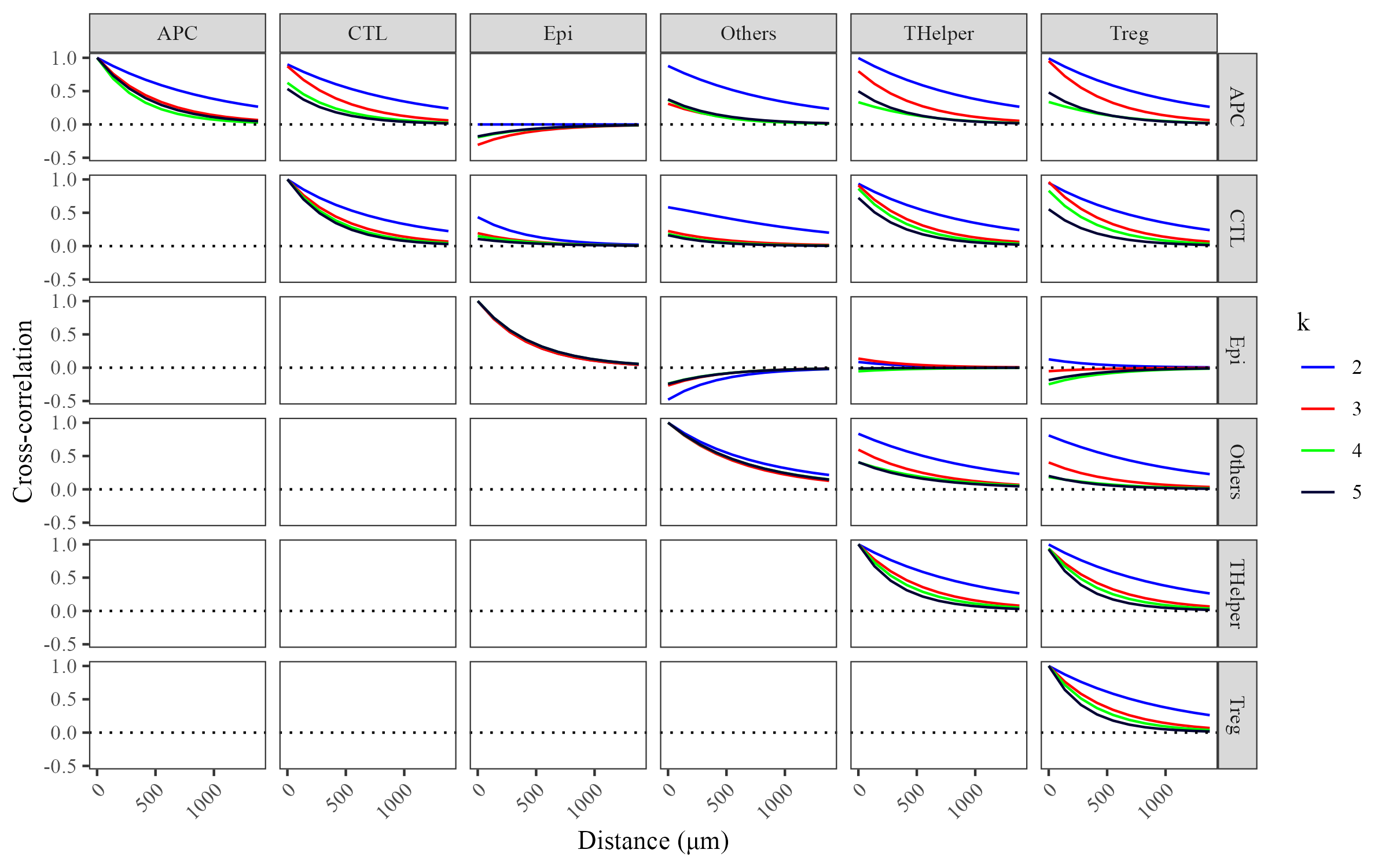}
\caption{Estimated cross-correlations at various distances, estimated at various $k$, in the IPMN patient group.}
\label{fig:panc_choosingk/xcor_varyingk_group2}
\vspace{1mm}
\end{figure}

Figures \ref{fig:panc_choosingk/xcor_varyingk_group1} and \ref{fig:panc_choosingk/xcor_varyingk_group2} presents the estimated cross-correlations at various spatial distances for different choices of $k$ in the PDAC and IPMN patient groups. The results indicate that cross-correlation estimates remain relatively stable for moderate to large values of $k$, whereas overly small values introduce larger deviations in the estimates.

\begin{table}[!h]
\centering
\caption{\label{tab:timings_panc}Model fitting times on images from pancreatic cancer patient groups for various $k$.}
\centering
\resizebox{\ifdim\width>\linewidth\linewidth\else\width\fi}{!}{
\begin{tabular}[t]{>{}ccc}
\toprule
\multicolumn{1}{c}{ } & \multicolumn{2}{c}{Model fitting time (s)} \\
\cmidrule(l{3pt}r{3pt}){2-3}
\textbf{k} & \textbf{PDAC} & \textbf{IPMN}\\
\midrule
\textbf{\cellcolor{gray!10}{2}} & \cellcolor{gray!10}{11934} & \cellcolor{gray!10}{7522}\\
\textbf{3} & 14852 & 9253\\
\textbf{\cellcolor{gray!10}{4}} & \cellcolor{gray!10}{18388} & \cellcolor{gray!10}{11307}\\
\textbf{5} & 21354 & 14258\\
\bottomrule
\end{tabular}}
\end{table}

\clearpage
\subsection{Analysis of PANC Dataset}

In this section, we analyze the cross-correlation differences between patient groups in the pancreatic cancer (PANC) dataset.

\begin{figure}[!htb]
\centering
\includegraphics[width=\textwidth]{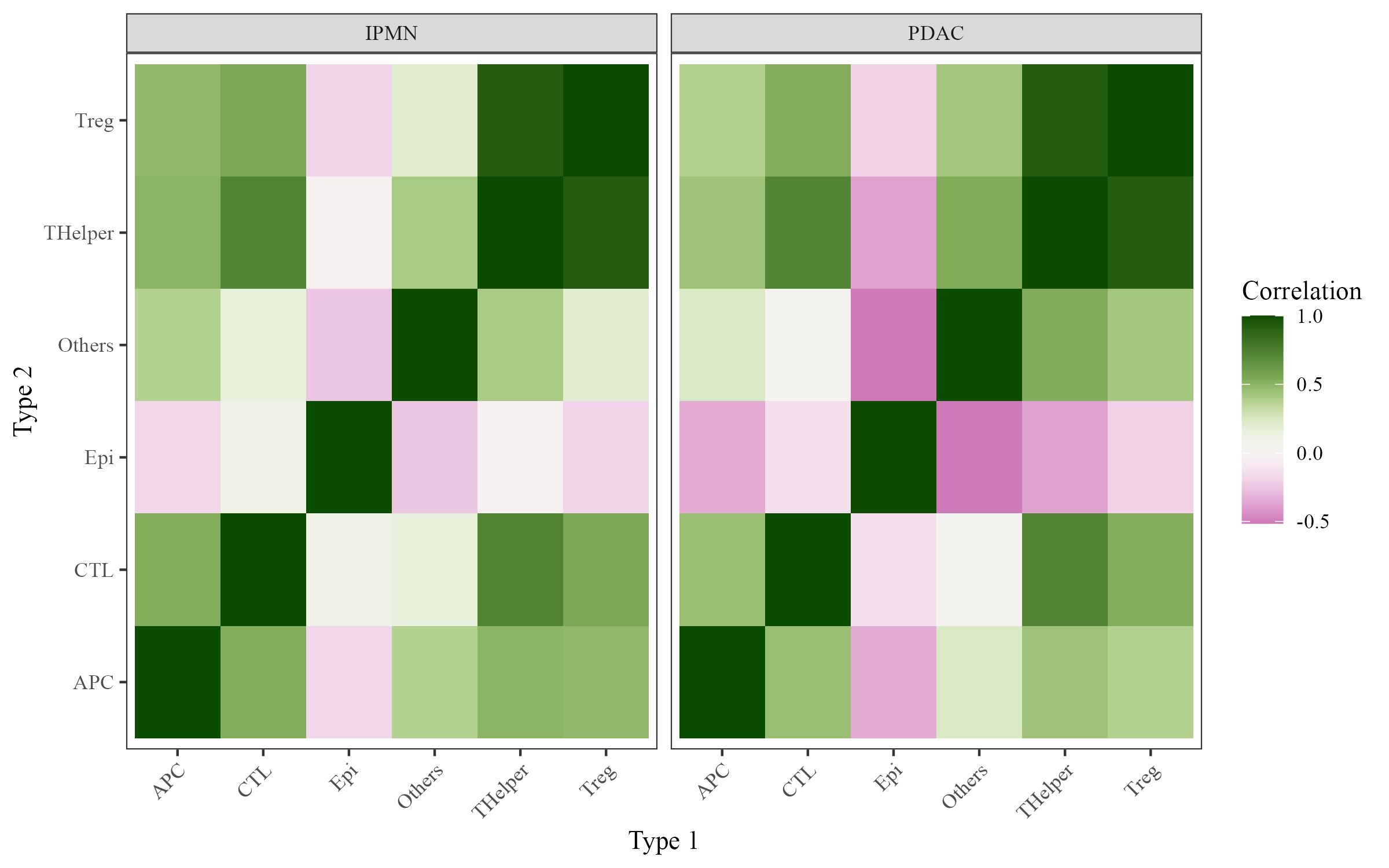}
\caption{Estimated posterior mean of the zero-distance ($h=0$) cross-correlation matrix between the different cell types in both patient groups from the pancreatic cancer dataset.}
\label{fig:exp_zero_cor_panc.png}
\vspace{1mm}
\end{figure}

Figure \ref{fig:exp_zero_cor_panc.png} shows the esimated mean of the zero-distance cross-correlations between each pair of cell types for both patient groups. One interesting thing to note is the exclusion of epithelial cells from all other cell types, in both groups.

\begin{figure}[!htb]
\centering
\includegraphics[width=\textwidth]{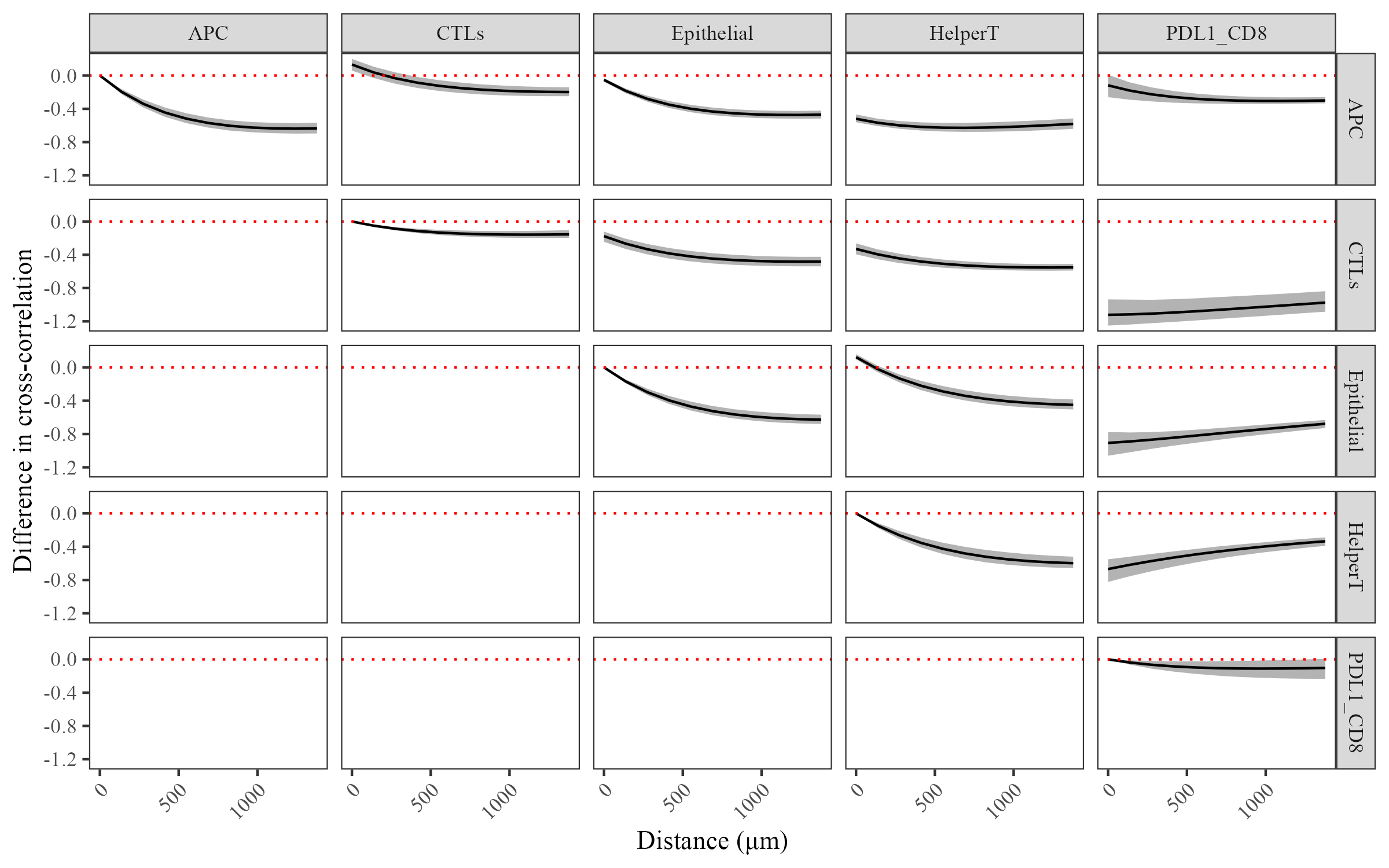}
\caption{The estimated difference in cross-correlation between patient groups for the select cell types. The black line in each plot indicates the posterior expectation, while the shaded region indicates the pointwise 95\% credible interval.}
\label{fig:diff_cor_panc}
\vspace{1mm}
\end{figure}

Figure \ref{fig:diff_cor_panc} presents the estimated difference in cross-correlation between the IPMN and PDAC patient groups for the same selected cell types as shown in the main article. These differences provide insight into how spatial cellular organization varies between disease subtypes, highlighting potential tumor-immune interactions that differentiate immune-permissive and immune-exclusive microenvironments.

\clearpage
\subsection{Choosing $k$ for CRC Analysis}

In this section, we assess the selection of the number of latent factors, $k$, for colorectal cancer (CRC) analysis by evaluating model convergence and the stability of cross-correlation estimates across two patient groups: CLR and DII.

Table \ref{tab:timings_crc} shows the model fitting times for various $k$ in both patient groups.

\begin{figure}[htb]
\centering
\includegraphics[width=\textwidth]{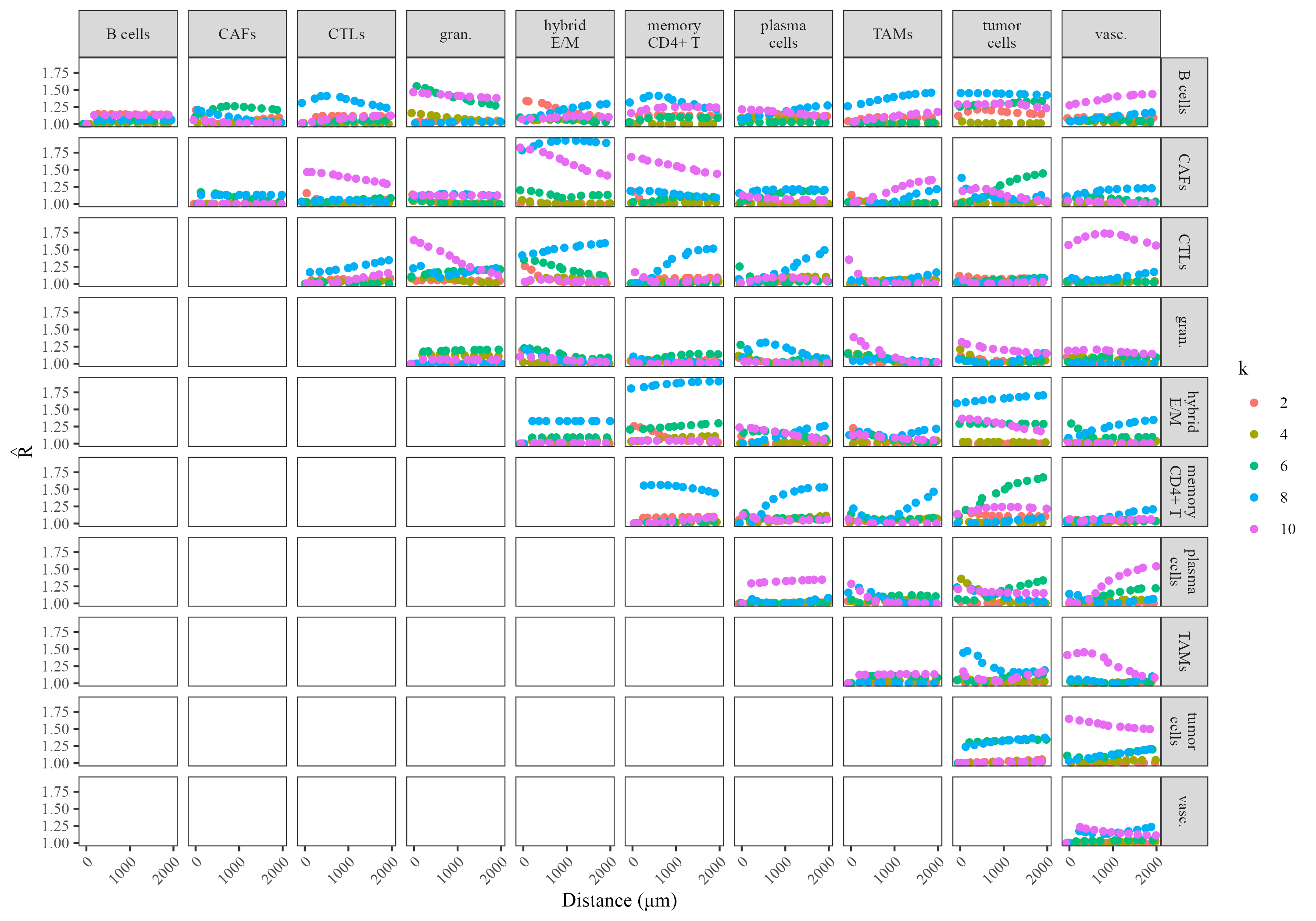}
\caption{$\hat{R}$ values to determine convergence of MCMC sampling of cross-correlations at various distances, under various $k$, in the CLR patient group. Values closer to 1 indicate better convergence.}
\label{fig:CRC_choosingk/rhat_1}
\vspace{1mm}
\end{figure}

\begin{figure}[htb]
\centering
\includegraphics[width=\textwidth]{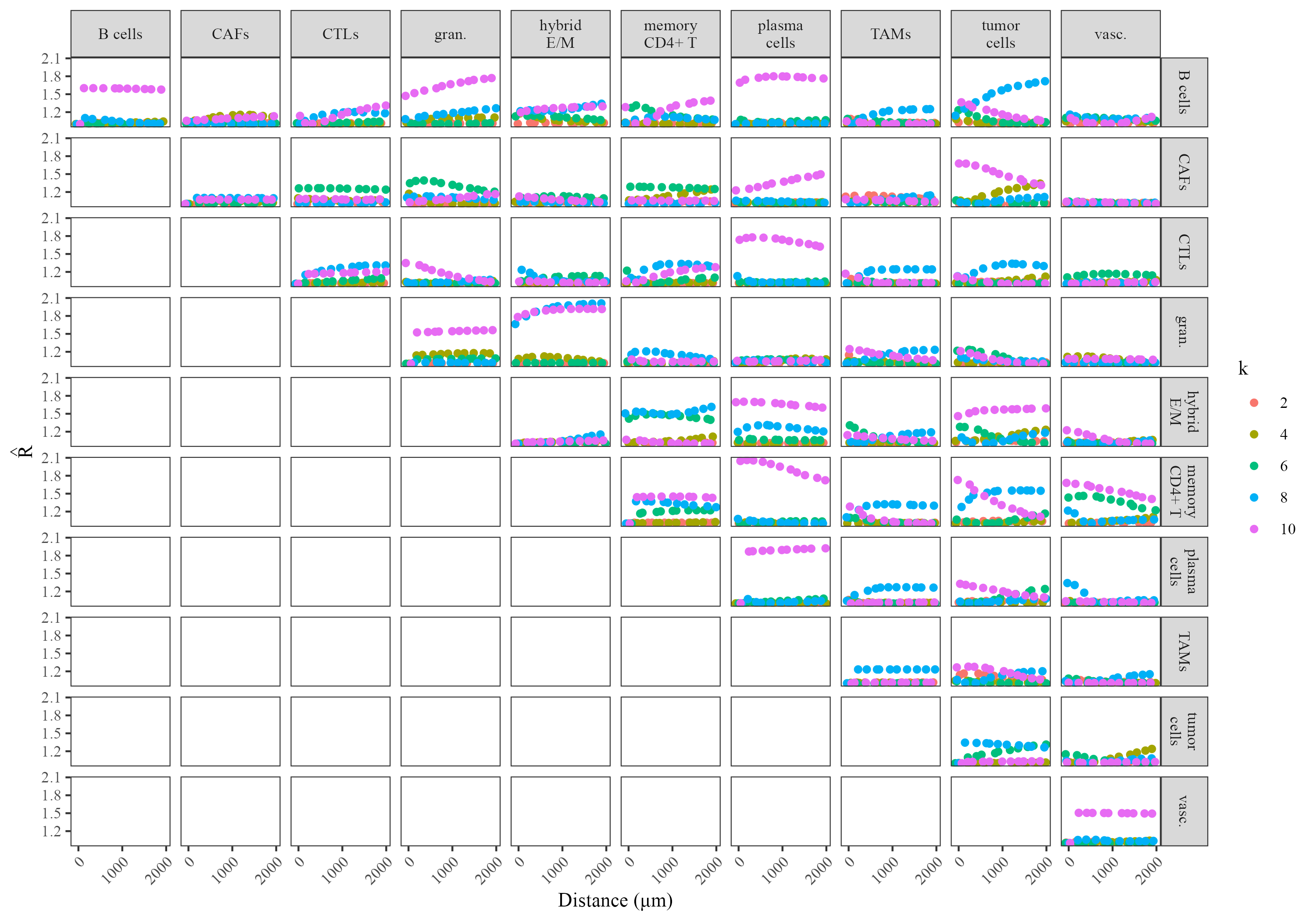}
\caption{$\hat{R}$ values to determine convergence of MCMC sampling of cross-correlations at various distances, under various $k$, in the DII patient group. Values closer to 1 indicate better convergence.}
\label{fig:CRC_choosingk/rhat_2}
\vspace{1mm}
\end{figure}

Figures \ref{fig:CRC_choosingk/rhat_1} and \ref{fig:CRC_choosingk/rhat_2} present the Gelman-Rubin diagnostic ($\hat{R}$) values for cross-correlation estimates at different spatial distances under varying $k$ in the CLR and DII patient groups, respectively. Values closer to 1 indicate better convergence, and the results suggest that low-to-moderate values of $k$ lead to more stable MCMC performance.

\begin{figure}[htb]
\centering
\includegraphics[width=\textwidth]{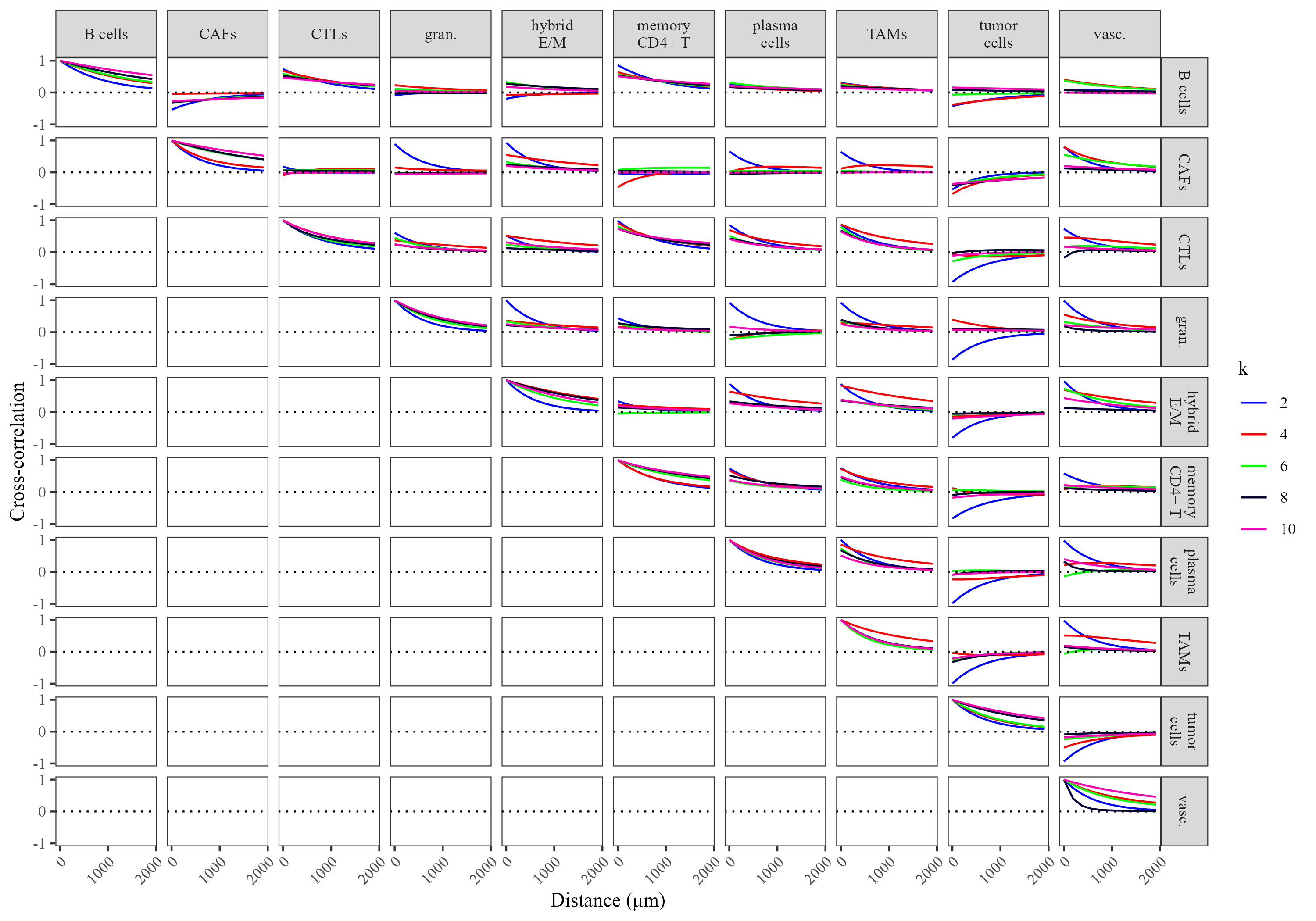}
\caption{Estimated cross-correlations at various distances, estimated at various $k$, in the CLR patient group.}
\label{fig:CRC_choosingk/xcor_varyingk_group1}
\vspace{1mm}
\end{figure}

\begin{figure}[htb]
\centering
\includegraphics[width=\textwidth]{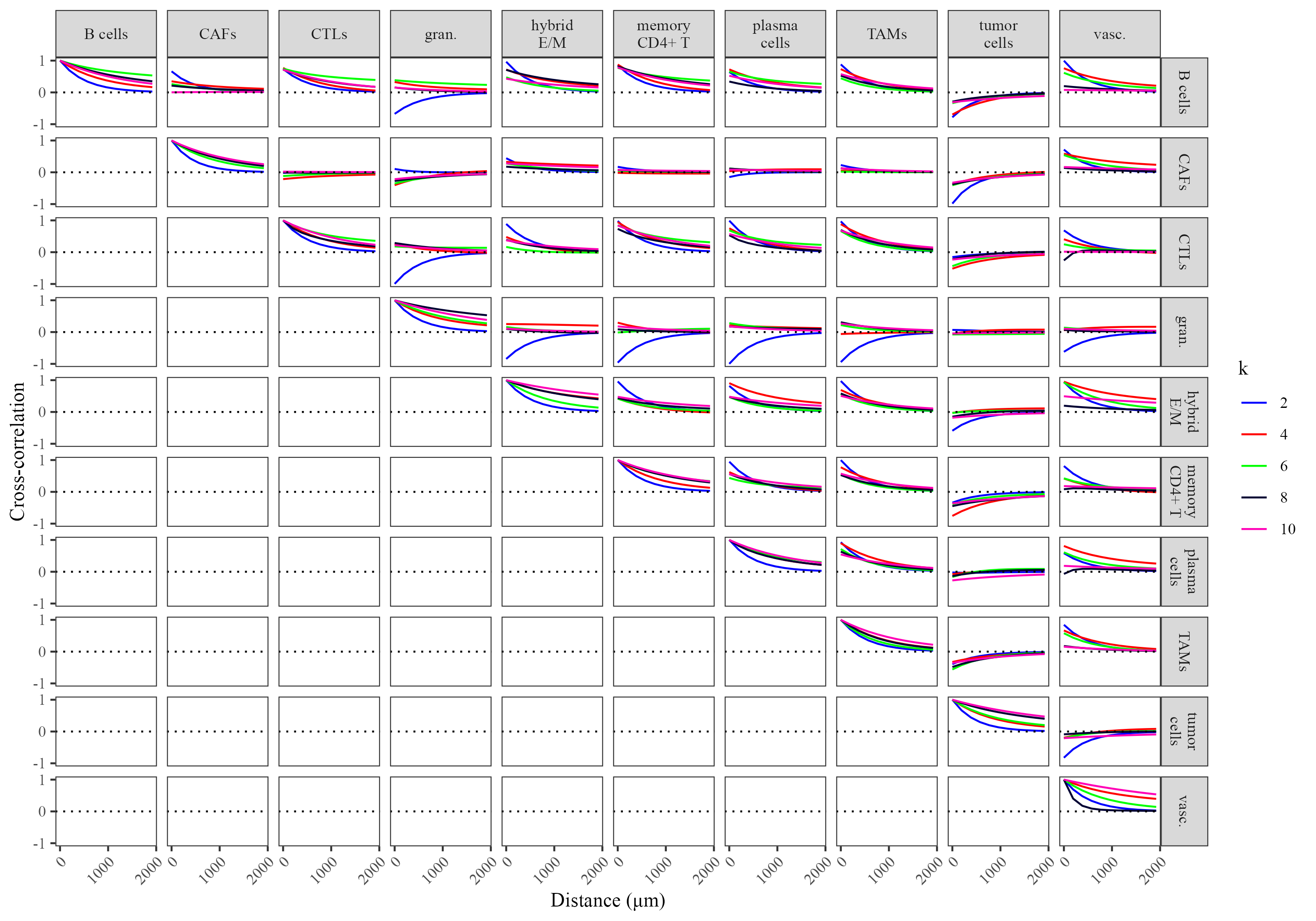}
\caption{Estimated cross-correlations at various distances, estimated at various $k$, in the DII patient group.}
\label{fig:CRC_choosingk/xcor_varyingk_group2}
\vspace{1mm}
\end{figure}

Figures \ref{fig:CRC_choosingk/xcor_varyingk_group1} and \ref{fig:CRC_choosingk/xcor_varyingk_group2} display estimated cross-correlations at different spatial distances for various $k$ values in the CLR and DII patient groups, respectively. These estimates demonstrate spatial cross-correlation curves estimated from models with low values of $k$ ($k=2,4$) have larger deviations, while those with $k\geq 6$ seem to have converged.

\begin{figure}[htb]
\centering
\includegraphics[width=\textwidth]{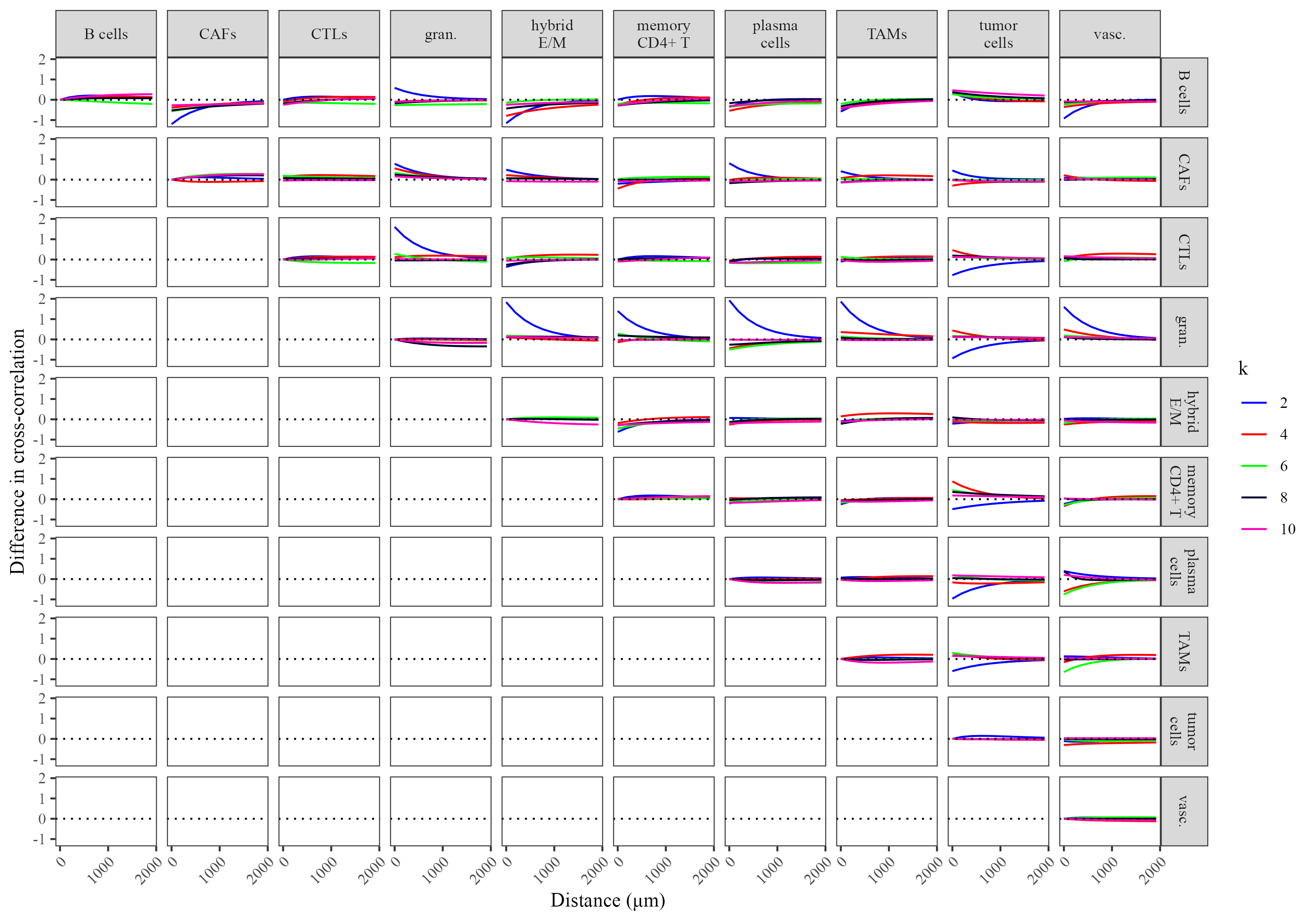}
\caption{Difference in estimated cross-correlations at various distances, estimated at various $k$, between the two patient groups.}
\label{fig:CRC_choosingk/diff_xcor_varyingk}
\vspace{1mm}
\end{figure}

Finally, Figure \ref{fig:CRC_choosingk/diff_xcor_varyingk} illustrates the differences in estimated cross-correlations between the two patient groups across varying $k$. Again, it seems that spatial cross-correlation curves estimated from models with low values of $k$ ($k=2,4$) have larger deviations, while those with $k\geq 6$ seem to have converged.

\begin{table}[!h]
\centering
\caption{\label{tab:timings_crc}Model fitting times on images from colorectal cancer patient groups for various $k$.}
\centering
\resizebox{\ifdim\width>\linewidth\linewidth\else\width\fi}{!}{
\begin{tabular}[t]{>{}ccc}
\toprule
\multicolumn{1}{c}{ } & \multicolumn{2}{c}{Model fitting time (s)} \\
\cmidrule(l{3pt}r{3pt}){2-3}
\textbf{k} & \textbf{CLR} & \textbf{DII}\\
\midrule
\textbf{\cellcolor{gray!10}{2}} & \cellcolor{gray!10}{10046} & \cellcolor{gray!10}{11410}\\
\textbf{4} & 14703 & 15898\\
\textbf{\cellcolor{gray!10}{6}} & \cellcolor{gray!10}{21533} & \cellcolor{gray!10}{23235}\\
\textbf{8} & 30255 & 32648\\
\textbf{\cellcolor{gray!10}{10}} & \cellcolor{gray!10}{43254} & \cellcolor{gray!10}{45180}\\
\bottomrule
\end{tabular}}
\end{table}

\clearpage
\subsection{Analysis of CRC Dataset}

In this section, we analyze spatial cross-correlations in the colorectal cancer (CRC) dataset, comparing cellular interactions between the CLR and DII patient groups.

\begin{figure}[!htb]
\centering
\includegraphics[width=\textwidth]{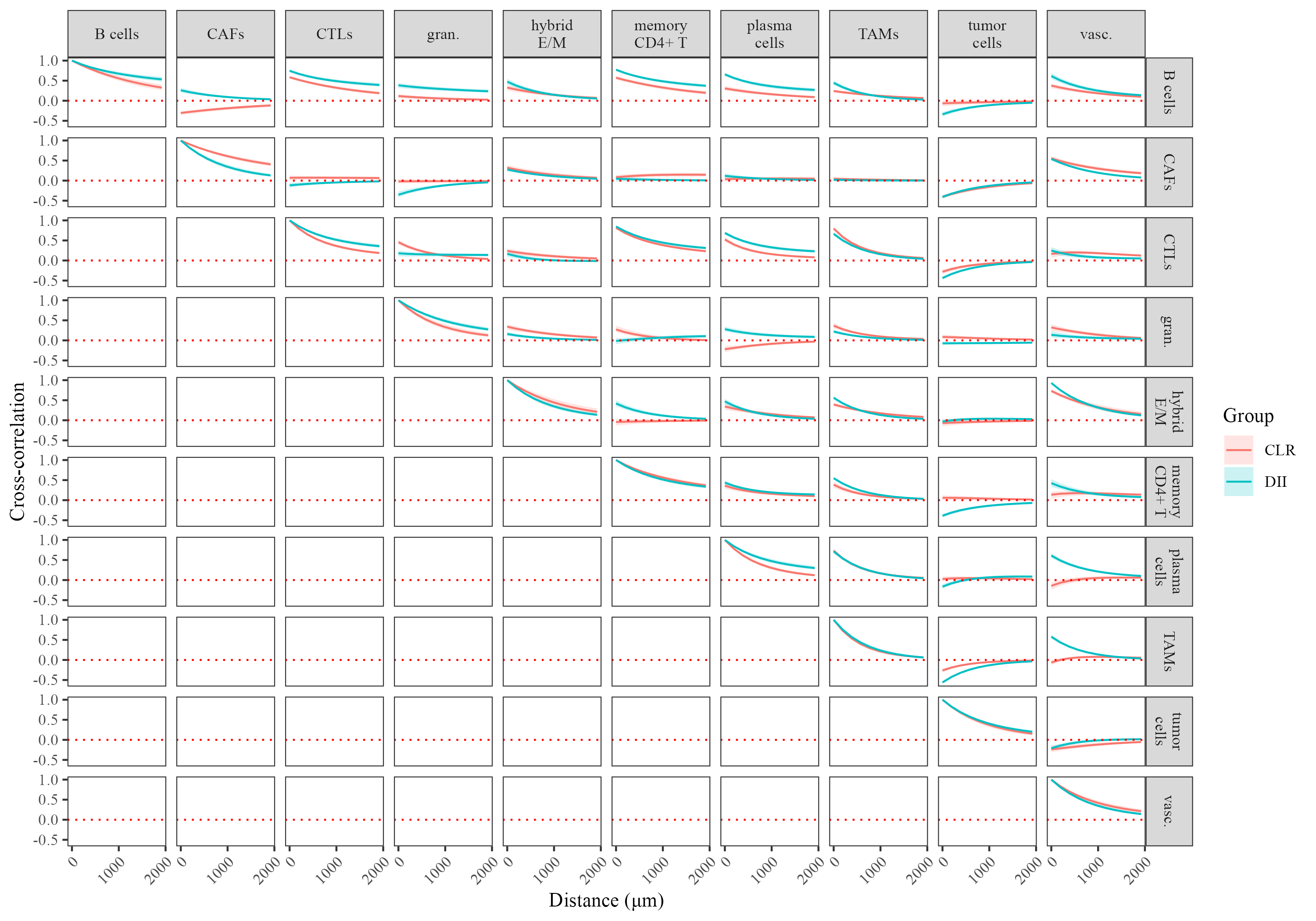}
\caption{The estimated spatial cross-correlations between select pairs of cell-type combinations from patients in the CLR group and the DII group. The colored line in each plot indicates the posterior expectation, while the shaded region indicates the pointwise 95\% credible interval.}
\label{fig:xcor_CRC_full}
\vspace{1mm}
\end{figure}

Figure \ref{fig:xcor_CRC_full} presents the estimated spatial cross-correlations for all cell-type combinations in both patient groups.

\begin{figure}[!htb]
\centering
\includegraphics[width=\textwidth]{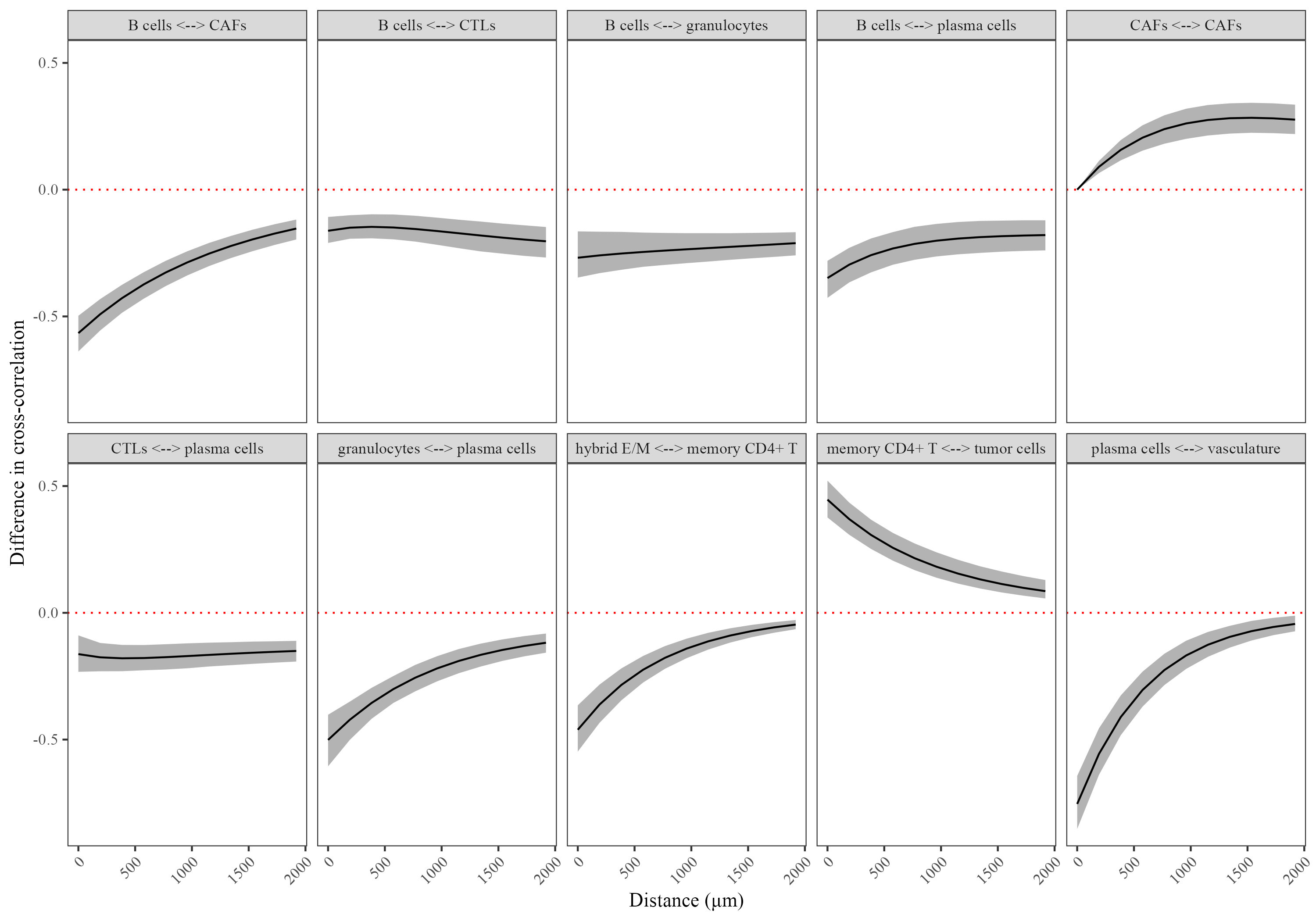}
\caption{The estimated difference in cross-correlation between patient groups for the same selected cell types as in the main article. The black line in each plot indicates the posterior expectation, while the shaded region indicates the pointwise 95\% credible interval.}
\label{fig:diff_cor_CRC}
\vspace{1mm}
\end{figure}

Figure \ref{fig:diff_cor_CRC} displays the estimated difference in cross-correlation between the two patient groups for the same selected cell types as in the main article. This comparison highlights key differences in spatial organization between the CLR and DII groups, potentially revealing distinct immune or stromal interactions associated with each subtype.

\clearpage

\section{Markov-chain Monte Carlo for posterior sampling}
We seek to sample from the posterior distribution of the following model 
\begin{equation}\label{appx:multi_subject_model}
\begin{aligned}
y_{ij}(\bl) \mid \lambda_{ij}(\bl) \sim \text{Poisson}(\lambda_{ij}(\bl))&, \qquad \blambda_i(\bl) = \begin{bmatrix} \lambda_{i1}(\bl) \cdots \lambda_{iq}(\bl) \end{bmatrix}^\top\\
\log\blambda_i(\bl) = \balpha_i + \bX_i(\bl)\bB + \bw_i(\bl)&, \qquad \bw_i(\bl)=\bA \bv_i(\bl),\\
\bv_i(\bl) = \begin{bmatrix} v_{i,1}(\bl), \dots, v_{i,k}(\bl) \end{bmatrix},&^\top \qquad v_{i,j}(\cdot) \sim GP(\bzero, \rho(\cdot,\cdot; \varphi_j)),
\end{aligned}
\end{equation}
where $y_{ij}(\bl)$ is the number of cells of type $j$ at location $\bl$ in the $i$th image. The MCMC algorithm we use for fitting our Bayesian hierarchical model is a straightforward multi-subject generalization of Algorithm 3 in \cite{peruzzi_spatial_2024}. The key differences are as follows:
\begin{itemize}
\item The joint update of the $j$th column of $\bB$ (labeled $\bB_{\cdot j}$) and $j$th row of $\bA$ (labeled $\bA_{j \cdot}$) in step 1 targets the full conditional posterior
\[N(\bF_j; \bzero, \bV_{F}) \prod_{i=1}^N \prod_{\bl} \text{Poisson}(y_{ij}(\bl); \lambda_{ij}(\bl)),  \]
where $\bF_j = (\bB_{\cdot j}^\top, \bA_{j \cdot})^\top$, we fix $\bV_F = \text{diag}\{ 10^3\}$, and we take advantage of conditional independence across subjects when performing the first product.
\item Because the Poisson density has no dispersion parameter, step 2 is not performed
\item Step 3 targets the density
\[ \prod_i N(\bv_i; \bzero, \tilde{\Cov_{\bvarphi}}),  \]
where $\bv_i$ is the vector that collects the spatial factors at all spatial locations for subject $i$ and $\tilde{\Cov_{\bvarphi}}$ is the covariance matrix induced by the spatial factor model using MGP processes. In practice, this density is efficiently evaluated as a product of small dimensional Gaussian densities, without any need to directly compute $\tilde{\Cov_{\bvarphi}}$ or its inverse. Refer to \cite{peruzzi_spatial_2024}, Section 2, for details on MGP models, and Section 4 for additional information on MGPs used in the context of spatial factor and coregionalization models such as those implemented here.
\item Step 4 is unchanged, but it is now enclosed in an additional loop that cycles through images and which can be performed in parallel. The updated step 4 thus samples the image-specific latent effects in parallel for all images in the data.
\end{itemize}

\section{Implementation details}
All the code to generate figures and simulations is available at \codereproduce. The R package (\rpack) is general purpose and fits our LGCP model with latent spatial factors. The R package is implemented in C++ using the Armadillo library \citep{armadillo} via RcppArmadillo \citep{rcpparmadillo}. Parallel operations are performed via OpenMP \citep{dagum1998openmp}. For fast linear algebra, we recommend to link R with efficient BLAS/LAPACK libraries \citep{lapack99} such as OpenBLAS \citep{openblas}. All real-data analyses were run on a workstation equipped with a 16-core AMD Ryzen 9 7950X3D CPU, 192GB memory, using Intel Math Kernel Library version 2024.2 \citep{intel2019mkl} for fast linear algebra. Simulation studies were run on an HPC platform - each simulation was run with 4 cores on a 3.0 GHz Intel Xeon Gold 6154 with 32GB of memory, again using the Intel Math Kernel Library version 2024.2.

\bibliographystyle{agsm}
\bibliography{references.bib}

\end{document}